\documentclass[12pt]{article}
\usepackage{graphicx,amssymb,amsmath,amscd,tensor,empheq,color}
\usepackage[colorlinks=true,urlcolor=blue,linkcolor=black,citecolor=black]{hyperref}

\makeatletter

\newlength{\fighskip} \fighskip=2pt
\newlength{\figvskip} \figvskip=3pt

\newcommand*{\figbox}[2]{{
  \def\figscale{#1}
  \def\arraystretch{0.8}
  \arraycolsep=0pt
  \begin{array}{c}
    \vbox{\vskip\figscale\figvskip
      \hbox{\hskip\figscale\fighskip
        \includegraphics[scale=\figscale]{#2}}}
  \end{array}}}

\newcommand*{\widebox}[1]{\setlength{\fboxsep}{1ex}%
  \fbox{#1}}
\newcommand*{\wideboxed}[1]{\setlength{\fboxsep}{1ex}%
  \fbox{\m@th$\displaystyle#1$}}

\newcommand*{\useshortskip}[1]{{%
\setlength\abovedisplayskip\abovedisplayshortskip#1}\ignorespaces}

\def\ubrace#1_#2{%
  \underbrace{#1}_{\hb@xt@\z@{\hss$\scriptstyle#2$\hss}}}

\newcommand{\hgf}{%
\,\tensor[_{2\kern-1.2pt}]{F}{_{\kern-0.8pt 1}}\kern-1.2pt}
\newcommand{\hgfs}{\mathbf{F}}

\newcommand{\blangle}{\bigl\langle}
\newcommand{\brangle}{\bigr\rangle}
\newcommand{\dlangle}{\langle\kern-1.5pt\langle}
\newcommand{\drangle}{\rangle\kern-1.5pt\rangle}
\newcommand{\bdlangle}{\blangle\kern-3pt\blangle}
\newcommand{\bdrangle}{\brangle\kern-3pt\brangle}

\newcommand*{\bra}[1]{\langle{#1}|}
\newcommand*{\ket}[1]{|{#1}\rangle}
\newcommand*{\braket}[2]{\langle{#1}|{#2}\rangle}

\newcommand*{\bbra}[1]{\blangle{#1}\big|}
\newcommand*{\bket}[1]{\big|{#1}\brangle}
\newcommand*{\bbraket}[2]{\blangle{#1}\big|{#2}\brangle}

\renewcommand{\le}{\leqslant}
\renewcommand{\ge}{\geqslant}

\newcommand{\ph}{\varphi}

\newcommand{\vth}{\vartheta}

\newcommand{\calC}{\mathcal{C}}
\newcommand{\calD}{\mathcal{D}}

\newcommand{\calF}{\mathcal{F}}

\newcommand{\calH}{\mathcal{H}}

\newcommand{\calM}{\mathcal{M}}

\newcommand{\calU}{\mathcal{U}}

\newcommand{\ZZ}{\mathbb{Z}}

\newcommand{\RR}{\mathbb{R}}
\newcommand{\CC}{\mathbb{C}}

\DeclareMathOperator{\sgn}{sgn}

\DeclareMathOperator{\diag}{diag}
\DeclareMathOperator{\Ld}{\mathcal{L}}

\DeclareMathOperator{\SO}{SO}

\DeclareMathOperator{\SU}{SU}
\DeclareMathOperator{\su}{\mathfrak{su}}

\DeclareMathOperator{\SL}{SL}
\DeclareMathOperator{\PSL}{PSL}
\DeclareMathOperator{\sL}{\mathfrak{sl}}

\DeclareMathOperator{\diff}{diff}
\DeclareMathOperator{\tDiff}{\widetilde{\mathrm{Diff}}}
\DeclareMathOperator{\AdS}{AdS}
\DeclareMathOperator{\HH}{H}

\let\Re\relax\DeclareMathOperator{\Re}{Re}
\let\Im\relax\DeclareMathOperator{\Im}{Im}

\newcommand{\spinup}{{\uparrow}}
\newcommand{\spindown}{{\downarrow}}

\newcommand{\la}{\text{L}}
\newcommand{\ra}{\text{R}}
\newcommand{\La}{\mathbf{L}}
\newcommand{\Ra}{\mathbf{R}}

\newcommand*{\matelem}[4]{{#1}_{#2,#4}^{\,#3}}
\newcommand{\oU}{\overline{U}\mkern-2mu}
\newcommand*{\Up}[3]{\matelem{U}{#1}{#2}{#3}}
\newcommand*{\Um}[3]{\matelem{\oU}{#1}{\mkern5mu#2}{#3}}
\newcommand*{\psie}[3]{\matelem{\psi}{#1}{#2}{#3}}

\newcommand{\GG}{\mathfrak{G}}
\newcommand{\tGG}{\widetilde{\GG}}
\newcommand{\tK}{\widetilde{K}}
\newcommand{\tf}{\tilde{f}}
\newcommand{\tg}{\tilde{g}}
\newcommand{\of}{\mathring{f}}

\newcommand{\arXiv}[1]{\href{http://arxiv.org/abs/#1}{\texttt{arXiv:#1}}}

\makeatother

\title{Notes on $\widetilde{\mathrm{SL}}(2,\mathbb{R})$ representations}
\author{Alexei Kitaev\\
\normalsize\it California Institute of Technology, Pasadena, CA 91125, U.S.A.}
\date{21 August 2018}

\begin{document}
\setcounter{tocdepth}{2}

\maketitle
\begin{abstract}
These notes describe representations of the universal cover of $\mathrm{SL}(2,\mathbb{R})$ with a view toward applications in physics. Spinors on the hyperbolic plane and the two-dimensional anti-de Sitter space are also discussed.
\end{abstract}

\tableofcontents
\newpage

\section*{Introduction}

The representations of the group $\SL(2,\RR)$ were originally studied by Bargmann~\cite{Bargmann47}, and an explicit Plancherel formula was obtained by Harish-Chandra~\cite{Harish-Chandra52}. The representation theory for the universal cover of that group is largely due to Puk\'anszky~\cite{Pukanszky64}. These notes are intended as an accessible (though not very rigorous) exposition, with notation and equations that would be convenient for general use. My main motivation comes from the study of the Sachdev-Ye-Kitaev (SYK) model~\cite{SY92,Kitaev-KITP,MS16}. This model has zero spatial dimensions, so the only coordinate is time $t$. At low temperatures, all correlation functions are invariant under M\"obius transformations $z\mapsto\frac{az+b}{cz+d}$ of the variable $z=\exp(2\pi t/\beta)$. If $t$ is real, then $z$ is also real, and thus, the symmetry group is $\PSL(2,\RR)=\SL(2,\RR)/\{1,-1\}$. It is, actually, more common to consider the problem in Euclidean time, $t=-i\tau$. In this case, the symmetry transformations preserve the unit circle (and also the orientation). The corresponding group $\GG$ is isomorphic to $\PSL(2,\RR)$. In general, one is interested in the action of $\GG$ on functions of two points on the circle. Fermionic wave functions (of one point on the circle) are transformed under the double cover of that group, i.e.\ $\SL(2,\RR)$. Furthermore, the study of the SYK model in Lorentzian time requires the universal cover of $\SL(2,\RR)$ (or $\GG$), which is denoted by $\tGG$.

Finite-dimensional representations of $\tGG$ are similar to those for $\SU(2)$: they are defined by an integer or half-integer spin $S$. However, these representations (except for the trivial one) are not unitary. All nontrivial unitary representations are infinitely dimensional. We will discuss three types of elementary representations:
\begin{enumerate}
\item Irreducible unitary representations, which include continuous and discrete series.
\item Forms of an arbitrary degree on the unit circle. These are, essentially, functions on the circle, whereas the degree $\lambda$ (which can be any complex number) enters the transformation law under diffeomorphisms.
\item Holomorphic $\lambda$-forms on the unit disk.
\end{enumerate}
The last two are also called ``non-unitary'' continuous and discrete series because they do not have a Hermitian inner product as part of their definition. However, they include the unitary irreps as special cases.

Unitary representations of $\tGG$ are sufficiently well-behaved; for example, an arbitrary unitary representation splits into irreps with multiplicities. (We do not prove this result because it involves a great deal of operator algebras.) Some differences from the more familiar case of compact Lie groups are as follows:
\begin{itemize}
\item Irreducible unitary representations can be infinitely-dimensional.
\item Some irreps do not enter the decomposition of the left regular representation. The irreps $\alpha$ in that decomposition are those whose matrix elements $\bra{l}U_\alpha(g)\ket{m}$ are normalizable or $\delta$-normalizable as functions of the group element $g$. The normalization factor is known as the \emph{Plancherel measure} on the set of irreps.
\end{itemize}

With physical applications in mind, we give explicit formulas for the matrix element functions and more general Casimir eigenfunctions. The latter may be interpreted as spinors on the anti-de Sitter space $\AdS_2$, or as functions (or forms) of two points on a circle.

\section{The group $\GG\cong\PSL(2,\RR)$ and its universal cover}

First, let us consider the the M\"obius group $\PSL(2,\CC)=\SL(2,\CC)/\{1,-1\}$. It consists of all linear fractional maps
\begin{equation}\label{Mobtrans}
z\mapsto \frac{az+b}{cz+d},\qquad
\text{where}\,\ a,b,c,d\in\CC,\quad ad-bc=1.
\end{equation}
The corresponding Lie algebra is denoted by $\sL_2(\CC)$. Its general element is a traceless matrix $\delta W=\begin{pmatrix}\delta a& \delta b\\ \delta c& \delta d\end{pmatrix}$, which generates an infinitesimal change of $z$ by $\delta z=\delta b +(\delta a-\delta d)z- (\delta c)z^2$. One can write $\delta W$ in a similar form, namely, $\delta W=(\delta b)L_{-1} +(\delta a-\delta d)L_0 -(\delta c)L_1$, where $\delta a+\delta d=0$ and
\begin{equation}\label{L2S}
L_{-1}=\begin{pmatrix} 0&1\\ 0&0\end{pmatrix},\qquad
L_{0}=\frac{1}{2}\begin{pmatrix} 1&0\\ 0&-1\end{pmatrix},\qquad
L_{1}=\begin{pmatrix} 0&0\\ -1&0\end{pmatrix}.
\end{equation}
The basis elements $L_n$ of the Lie algebra may be interpreted as complex vector fields with the following $z$-components and commutation relations:\footnote{We define the action of a vector field $v$ on functions by the formula $f\mapsto -\partial_{v}f=-v^\beta\partial_{\beta}f$. Correspondingly, $[u,v]=-\Ld_{u}v$, where $\Ld$ denotes the Lie derivative.}
\begin{equation}
L_n^z(z)=z^{n+1},\qquad [L_n,L_m]=(n-m)L_{n+m}.
\end{equation}

Now, the subgroup $\PSL(2,\RR)$ consists of the M\"obius transformations with real parameters $a,b,c,d$. Such maps preserve the upper half-plane. However, it is more convenient to work with the subgroup $\GG$ of those transformations that preserve the unit disk $|z|<1$. It is conjugate to $\PSL(2,\RR)$ by the Cayley map $z\mapsto\frac{z-i}{z+i}$, which takes the upper half-plane to the unit disk. A standard basis of the corresponding Lie algebra $\mathfrak{g}$ is as follows:
\begin{equation}\label{SL2gen}
\begin{alignedat}{3}
\Lambda_0 &=\frac{1}{2}\begin{pmatrix}i&0\\ 0&-i\end{pmatrix}
&&=\quad\: iL_0\qquad\quad
&&\figbox{1.0}{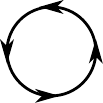} \\[5pt]
\Lambda_1 &=\frac{1}{2}\begin{pmatrix}0&1\\  1&0\end{pmatrix}
&&=\frac{L_{-1}-L_{1}}{2}\qquad\quad
&&\figbox{1.0}{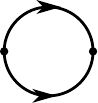} \\[5pt]
\Lambda_2 &=\frac{1}{2}\begin{pmatrix}0&i\\ -i&0\end{pmatrix}
&&=\frac{iL_{-1}+iL_{1}}{2}\qquad\quad
&&\figbox{1.0}{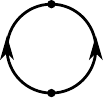}
\end{alignedat}
\end{equation}
One can also define $\mathfrak{g}$ abstractly, using the Lie bracket:
\begin{equation}
[\Lambda_0,\Lambda_1]=\Lambda_2,\qquad\quad
[\Lambda_0,\Lambda_2]=-\Lambda_1,\qquad\quad
[\Lambda_1,\Lambda_2]=-\Lambda_0.
\end{equation}
Up to a constant factor, the Casimir operator is
\begin{equation}\label{Casimir}
Q=\Lambda_0^2-\Lambda_1^2-\Lambda_2^2
=-L_0^2+\frac{1}{2}(L_{-1}L_1+L_1L_{-1}).
\end{equation}

The Lie algebra $\mathfrak{g}$ canonically defines the simply connected Lie group $\tGG$, the universal cover of $\GG$. Conversely, $\GG$ is the quotient of $\tGG$ by its center, $Z=\{e^{2\pi n\Lambda_0}:\,n\in\ZZ\}$.

\section{Irreducible unitary representations}

An irreducible representation of $\tGG$ is characterized by the eigenvalues of the Casimir operator $Q$ and the central element $e^{2\pi \Lambda_0}$. The latter has the form $e^{-2\pi i\mu}$, where $\mu\in\RR/\ZZ$. Representations of $\GG$ (rather than its universal cover) are characterized by $e^{-2\pi i\mu}=1$. By unitarity, $\Lambda_0$, $\Lambda_1$, $\Lambda_2$ are represented by anti-Hermitian operators; hence the Casimir eigenvalue $q$ is real. It and can be parametrized in a way similar to the expression $S(S+1)$ in the $\su(2)$ case:
\begin{equation}
q=\lambda(1-\lambda),\qquad
\text{where}\quad
\lambda\in\RR\quad \text{or}\quad 
\lambda=\tfrac{1}{2}+is,\: s\in\RR.
\end{equation}

The basis vectors $\ket{m}$ are eigenvectors of $\Lambda_0$ with the eigenvalues $-im$, where $m\in\mu+\ZZ$. Recall that $\Lambda_0=iL_0$ (see Eq.~(\ref{SL2gen})); hence $L_0\ket{m}=-m\ket{m}$. The commutation relations $[L_0,L_{-1}]=L_{-1}$ and $[L_0,L_{1}]=-L_{1}$ imply that $L_{-1}$ lowers, and $L_{1}$ raises, $m$ by $1$. Using the relation $[L_1,L_{-1}]=2L_{0}$ and the expression for the Casimir operator, $Q=-L_0^2 +\frac{1}{2}(L_{-1}L_1+L_1L_{-1})$, we find that
\begin{align}
\label{irrme1}
\bra{m}L_{1}L_{-1}\ket{m}&=m(m-1)+q=(m-\lambda)(m-1+\lambda),\\[3pt]
\label{irrme2}
\bra{m}L_{-1}L_{1}\ket{m}&=m(m+1)+q=(m+\lambda)(m+1-\lambda).
\end{align}
Since $L_n^\dag=L_{-n}$ due to unitarity, the matrix elements in the above equations are nonnegative for all values of $m$ that occur in a given representation. This condition is satisfied by three types of irreducible representations:
\begin{equation}
\wideboxed{
\begin{array}{@{\bullet\:\,}l@{\qquad\:}c@{\qquad}l}
\text{Trivial representation } I: & \mu=q=0, & m=0;\\[5pt]
\text{Continuous series } \calC^\mu_q:
& \begin{array}{@{}c@{}}
q>|\mu|(1-|\mu|),\\ \text{where } |\mu|\le 1/2,
\end{array}
& m=\mu+k\quad (k\in\ZZ);\\[5pt]
\text{Discrete series } \calD^+_\lambda,\,\calD^-_\lambda:
& \lambda>0,\:\: \mu=\pm\lambda,
& \begin{array}{@{}l@{}}
m=\lambda,\,\lambda+1,\,\lambda+2,\ldots \text{ or}\\
m=-\lambda,\,-\lambda-1,\,-\lambda-2,\ldots\,.
\end{array}
\end{array}
}
\end{equation}
We will also use the generic notation $\calU^\mu_\lambda$ that encompasses all three cases. The continuous series $\calC^\mu_q$ is further subdivided into the \emph{principal series,} $q\ge\frac{1}{4}$ (i.e.\ $\lambda=\frac{1}{2}+is$) and \emph{complementary series,} $q<\frac{1}{4}$  (i.e.\ $|\mu|<\lambda<\frac{1}{2}$).

The general solution to equations~(\ref{irrme1}), (\ref{irrme2}) involves some arbitrary phase factors, which can be absorbed in the definition of the basis vectors. Thus,
\begin{equation}
\label{L_irrep}
\wideboxed{
\begin{aligned}
L_{-1}\ket{m}&=-\sqrt{(m-\lambda)(m-1+\lambda)}\,\ket{m-1},\\[3pt]
L_0\ket{m}&=-m\,\ket{m},\\[3pt]
L_1\ket{m}&=-\sqrt{(m+\lambda)(m+1-\lambda)}\,\ket{m+1}.
\end{aligned}
}
\end{equation}


\section{Non-unitary continuous and discrete series}\label{sec_forms}

\subsection{The individual representations}

Let $\lambda\in\CC$ and $\mu\in\CC/\ZZ$. The non-unitary continuous series representation $\calF^\mu_\lambda$ consists of ``$\mu$-twisted $\lambda$-forms'' on the unit circle, which may be written as $f=\tf(\ph)\,(d\ph)^{\lambda}$. More formally, an abstract form $f$ is represented by a function $\tf$ of a real variable that satisfies a twisted periodicity condition and transforms under diffeomorphisms in a particular way:
\begin{empheq}[box=\widebox]{gather}
\tf(\ph+2\pi)=e^{2\pi i\mu}\tf(\ph)
\\[3pt]
\label{mapact_lambda}
(V\tf)(\ph)
=\bigl(\partial_{\ph}V^{-1}(\ph)\bigr)^\lambda\, \tf(V^{-1}(\ph))
\end{empheq}
Here $V$ is an element of the group $\tDiff_{+}(S^1)$, that is, a smooth monotone map $V:\,\RR\to\RR$ such that $V(\ph+2\pi)=V(\ph)+2\pi$. Infinitesimal diffeomorphisms, i.e.\ vector fields, act as follows:
\begin{equation}\label{vecact_lambda}
(v\tf)(\ph)
=-v^{\ph}\,\partial_{\ph}\tf(\ph)-\lambda\,(\partial_{\ph}v^{\ph})\,\tf(\ph).
\end{equation}
For example, if $\lambda=-1$, then $f$ is a vector field,  and $vf$ is equal to $-\Ld_{v}f$. One can also characterize the $\diff(S^1)$ action by applying Eq.~(\ref{vecact_lambda}) to the complex vector fields $L_n$ (expressed as $L_n^\ph=(\partial z/\partial\ph)^{-1}L_n^z=-ie^{in\ph}$) and the Fourier basis of the space of forms:
\begin{equation}\label{Lnfm}
\wideboxed{
L_{n}f_{\lambda,m}=-(m+n\lambda)f_{\lambda,m+n},\quad\: \text{where }\, \tf_{\lambda,m}(\ph)=e^{im\ph}\quad\: \text{for }\, m\in\mu+\ZZ
}
\end{equation}

To add some extra rigor, the elements of $\calF^\mu_\nu$ are required to have finite Sobolev norm of degree $d=\frac{1}{2}-\Re\lambda$. By definition, the norm of $f=\sum_{m}a_{m}f_{\lambda,m}$ is $\|f\|=\sqrt{\sum_{m}(1+m^2)^d|a_m|^2}$. If $\lambda<0$, then  $f$ is continuous; the larger the $\lambda$, the more singular can such functions be. While the norm remains finite under smooth diffeomorphisms, it is not preserved. There is, however, a nondegenerate $\tDiff_{+}(S^1)$-invariant pairing
\begin{equation}
\label{pairing1}
(g,f)=\int_{0}^{2\pi}\tg(\ph)\,\tf(\ph)\,\frac{d\ph}{2\pi}\,,\qquad
\text{where}\quad
g\in\calF^{-\mu}_{1-\lambda},\quad f\in\calF^\mu_\lambda.
\end{equation}
If $\mu$ is real and $\lambda=\frac{1}{2}+is$ for $s\in\RR$,\, then the formula $\braket{g}{f}=(g^*,f)$ defines a Hermitian inner product on $\calF^\mu_\lambda$.

We now regard $\calF^\mu_\lambda$ as a representation of $\tGG\subseteq\tDiff_{+}(S^1)$, i.e.\ specialize Eq.~(\ref{Lnfm}) to $n=-1,0,1$. As expected, the central element $e^{2\pi iL_0}$ and the Casimir operator $Q$ are represented by $e^{-2\pi i\mu}$ and $\lambda(1-\lambda)$, respectively. Thus, the current use of $\mu$ and $\lambda$ is consistent with that for unitary irreps.\smallskip

The representation $\calF_\lambda^\mu$ of the group $\tGG$ can be reducible. For example $\calF^{1/2}_{1/2}$ splits into the subspaces $\calF^{-}_{1/2}$ and $\calF^{+}_{1/2}$ that are spanned by the basis vectors $f_{1/2,m}$ with negative and positive $m$, respectively. In some other cases, the representation $\calF^\mu_\lambda$ has an invariant subspace but does not split. For example, let $\mu=\lambda\not=\frac{1}{2}$. Then $L_{1}f_{\lambda,\lambda-1}=af_{\lambda,\lambda}$, where $a=1-2\lambda\not=0$. However, $L_{-1}f_{\lambda,\lambda}=0$. This situation is indicated by the arrow in the diagram
\[
\figbox{1.0}{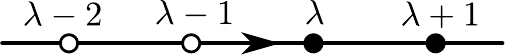}
\]
where the small circles represent the basis vectors $f_{\lambda,m}$ labeled by $m$. A line between $m-1$ and $m$ without an arrow means that both $L_{1}f_{\lambda,m-1}$ and $L_{-1}f_{\lambda,m}$ are nonzero. The vectors $f_{\lambda,\lambda},f_{\lambda,\lambda+1},\ldots$ (represented by the full circles) form a $\tGG$-invariant subspace $\calF_\lambda^+$, which has no invariant complement. Indeed, suppose that such a complement $(\calF_\lambda^+)^\perp$ exists. Any invariant subspace has a homogeneous basis, that is, the basis vectors are eigenvectors of $L_0$. Thus, $(\calF_\lambda^+)^\perp$ should be equal to the linear span of $f_{\lambda,\lambda-1},f_{\lambda,\lambda-2},\ldots$ (the empty circles). But this subspace is not, actually, invariant because $L_{1}f_{\lambda,\lambda-1}$ does not belong to it.

By definition, the ``non-unitary discrete series'' representations are the invariant subspaces $\calF_\lambda^+\subseteq\calF_\lambda^{\lambda}$ and $\calF_\lambda^-\subseteq\calF_\lambda^{-\lambda}$ that are spanned by basis vectors $f_{\lambda,m}$ with $m=\lambda,\lambda+1,\ldots$ in the first case and $m=-\lambda,-\lambda-1,\ldots$ in the second case. Elements of $\calF_\lambda^+$ and $\calF_\lambda^-$ may be written as
\begin{equation}
f=\of(z)\,(-i\,dz)^{\lambda},\quad\:
\of(e^{i\ph})=e^{-i\lambda\ph}\tf(\ph)
\qquad \text{or}\qquad
f=\of(z)\,(i\,dz^{-1})^{\lambda},\quad\:
\of(e^{i\ph})=e^{i\lambda\ph}\tf(\ph),
\end{equation}
respectively, where $\of(z)$ is holomorphic in the disk $|z|<1$ or its complementary domain on the Riemann sphere, $|z^{-1}|<1$. In this notation, the transformation rule~(\ref{mapact_lambda}) becomes
\begin{equation}\label{aftrans}
\wideboxed{
(V\of)(z)=\biggl(\frac{dz}{dw}\biggr)^{\!-\lambda}\of(w)
\quad\text{if } f\in\calF_\lambda^+,\qquad
(V\of)(z)=\biggl(\frac{dz^{-1}}{dw^{-1}}\biggr)^{\!-\lambda}\of(w)
\quad\text{if } f\in\calF_\lambda^-
}
\end{equation}
where $z=V(w)$ and $V\in\tGG$.

\subsection{Isomorphisms and intertwiners}

\begin{figure}
\centering\(
\calF^{1/2}_{1/2}:\: \figbox{1.0}{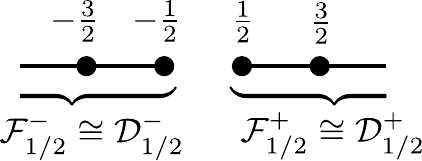}
\hspace{2cm}
\begin{array}{@{}c@{\:}c@{}}
\calF^{\lambda}_{\lambda}: & \figbox{1.0}{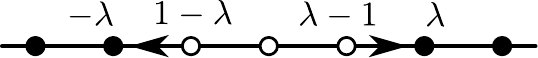}\\[10pt]
\calF^{\lambda}_{1-\lambda}: & \figbox{1.0}{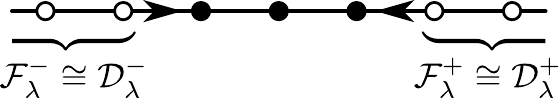}\\[10pt]
& \lambda=1,\frac{3}{2},2,\ldots
\end{array}
\)
\caption{The structure of representations $\calF^\mu_\lambda$ for some $\lambda$ and $\mu$. The full circles indicate those basis vectors $f_{\lambda,m}$ that span an invariant subspace, whereas the empty circles correspond to quotients.}
\label{fig_formrep}
\end{figure}

The structure of representations $\calF_\lambda^\mu$, $\calF_\lambda^{\pm}$, and their relations can be described as follows (see also Figure~\ref{fig_formrep}):
\begin{enumerate}
\item If $\lambda,-\lambda\notin\mu+\ZZ$, then $\calF_\lambda^\mu$ is irreducible and isomorphic to $\calF_{1-\lambda}^\mu$.
\item Let\, $\lambda\in\mu+\ZZ$\, or\, $-\lambda\in\mu+\ZZ$.\, (Without loss of generality, we may assume that $\mu=\pm\lambda$.) If $\lambda\notin\bigl\{0,-\tfrac{1}{2},-1,-\tfrac{3}{2},\ldots\bigr\}$, then the representation $\calF_\lambda^\pm\subseteq\calF_\lambda^{\mu}$ is irreducible and isomorphic to $\calF_{1-\lambda}^{\mu}/\calF_{1-\lambda}^\mp$.
\item If the unitary representation $\calC_{\lambda(1-\lambda)}^{\mu}$ or $\calD_{\lambda}^{\pm}$ exists for given $\lambda$ and $\mu$, it is isomorphic (except for the inner product) to $\calF^{\mu}_{\lambda}\cong\calF^{\mu}_{1-\lambda}$ or $\calF_{\lambda}^{\pm} \cong\calF_{1-\lambda}^{\pm\lambda}/\calF_{1-\lambda}^\mp$, respectively.
\end{enumerate}

The statements about irreducibility follow directly from Eq.~(\ref{Lnfm}). The isomorphisms in~1 and~2 can be obtained in a unified fashion using the intertwiners
\begin{equation}
\Xi_\lambda^{\mu\pm}:\, \calF_{1-\lambda}^\mu\to\calF_\lambda^\mu,\qquad\quad
\Xi_\lambda^{\mu\pm}f_{1-\lambda,m}=b_{\lambda,m}^{\pm}f_{\lambda,m}\quad\:
\text{for }\, m\in\mu+\ZZ,
\end{equation}
where
\begin{gather}
b_{\lambda,m}^{+}=\frac{\Gamma(\lambda+m)}{\Gamma(1-\lambda+m)}\quad\:
\text{if }\, \lambda+m\notin\ZZ,\qquad\quad
b_{\lambda,m}^{-}=\frac{\Gamma(\lambda-m)}{\Gamma(1-\lambda-m)}\quad\:
\text{if }\, \lambda-m\notin\ZZ;
\displaybreak[0]\\[8pt]
b_{\lambda,\lambda+k}^{+}=b_{\lambda,-(\lambda+k)}^{-}
=\begin{cases}
\displaystyle \frac{\Gamma(2\lambda+k)}{k!}
&\text{for } k=0,1,2,\ldots\\[3pt]
0 & \text{for } k=-1,-2.\ldots
\end{cases}\qquad
\text{if }\lambda\notin\bigl\{0,-\tfrac{1}{2},-1,\ldots\bigr\}.
\end{gather}
The above definitions agree when they overlap. One can check that $\Xi_\lambda^{\mu,\pm}$ is an intertwiner and that
\begin{gather}
\Xi_\lambda^{\mu,-}
=\frac{\sin\pi(\lambda+\mu)}{\sin\pi(\lambda-\mu)}\,\Xi_\lambda^{\mu,+},\qquad
\Xi_{1-\lambda}^{\mu,\pm}=\bigl(\Xi_\lambda^{\mu,\pm}\bigr)^{-1}\qquad
\text{if\, $\pm\lambda\notin\mu+\ZZ$};
\displaybreak[0]\\[5pt]
\operatorname{Kernel}(\Xi_\lambda^{\pm\lambda,\pm})
=\calF_{1-\lambda}^{\mp},\qquad
\operatorname{Image}(\Xi_\lambda^{\pm\lambda,\pm})
=\calF_{\lambda}^{\pm}.
\end{gather}
Thus, $\Xi_\lambda^{\mu,\pm}$ defines the required isomorphism (either between the full spaces or between the subspace and the quotient). Furthermore, $\Xi_\lambda^{\mu,\pm}$ and $\Xi_\lambda^{-\mu,\mp}$ are conjugate to each other:
\begin{equation}
\bigl(g,\,\Xi_\lambda^{\mu,\pm}f\bigr)
=\bigl(\Xi_\lambda^{-\mu,\mp}g,\,f\bigr).
\end{equation}
The maps $\Xi_\lambda^{\mu\pm}$ are bounded with respect to the Sobolev norm. In fact, $\bigl\|\Xi_\lambda^{\mu+}f_{1-\lambda,m}\bigr\| \approx\bigl\|f_{1-\lambda,m}\bigr\|$ for large positive $m$ and $\bigl\|\Xi_\lambda^{\mu-}f_{1-\lambda,m}\bigr\| \approx\bigl\|f_{1-\lambda,m}\bigr\|$ for large negative $m$.\smallskip

Now, let $\calU^\mu_\lambda$ be a nontrivial unitary irrep, i.e.\ either $\calC_{\lambda(1-\lambda)}^{\mu}$ or $\calD_{\lambda}^{\pm}$. Its isomorphism with $\calF_{\lambda}^{\mu}$ or $\calF_{\lambda}^{\pm}$ is obtained by factoring $\Xi_\lambda^{\mu\pm}$ as follows:
\begin{equation}\label{Xiupdown}
\wideboxed{
\calF_{1-\lambda}^\mu
\xrightarrow{\:\Xi_{\lambda\spindown\:}^{\mu\pm}}
\calU^\mu_\lambda \xrightarrow{\:\Xi_{\lambda\spinup}^{\mu\pm}\:}
\calF_{\lambda}^{\mu},\qquad\qquad
\Xi_{\lambda\spindown}^{\mu\pm}f_{1-\lambda,m}=c^{\pm}_{\lambda,m}\ket{m},
\qquad
\Xi_{\lambda\spinup}^{\mu\pm}\ket{m}=c^{\pm}_{\lambda,m}f_{\lambda,m}
}
\end{equation}
such that the first map is onto and the second is injective. The coefficients $c^{\pm}_{\lambda,m}$ are given by these formulas:
\begin{equation}\label{clamk}
c^{+}_{\lambda,m}
=\sqrt{\frac{\Gamma(\lambda+m)}{\Gamma(1-\lambda+m)}},\qquad\quad
c^{-}_{\lambda,m}
=(-1)^{m-\mu}\sqrt{\frac{\Gamma(\lambda-m)}{\Gamma(1-\lambda-m)}}.
\end{equation}
For the continuous series, they are both well-defined and their ratio is a function of $\lambda$ and $\mu$, namely $c^{+}_{\lambda,m}/c^{-}_{\lambda,m} =\sqrt{\sin\pi(\lambda-\mu)/\sin\pi(\lambda+\mu)}$. For the discrete series $\calD^+_\lambda$ (or $\calD^-_\lambda$), only $c^{+}_{\lambda,m}$ (resp.\ $c^{-}_{\lambda,m}$) exists.

The signs of the square roots in Eq.~(\ref{clamk}) require some care. Let us fix them on a case-by-case basis. For the continuous series, we set $c^{\pm}_{1/2,0}=1$ and analytically continue the functions $c^{\pm}_{\lambda,m}$ to $\lambda=\frac{1}{2}+is$ while keeping $m$ equal to $0$. If $s\not=0$, we further continue to all real $m$. If $s=0$ but $|\mu|<\frac{1}{2}$, then we can take the limit $c^{\pm}_{1/2,m}=\lim_{s\to 0}c^{\pm}_{1/2+is,m}$ to obtain the following expressions:
\begin{equation}
c^{+}_{1/2,m}=c^{-}_{1/2,m}=\gamma_{m-\mu},\qquad\quad
\text{where}\quad
\gamma_k= \begin{cases}
1 & \text{if } k\ge 0,\\
(-1)^k & \text{if } k<0.
\end{cases}
\end{equation}
For the complementary series, the correct signs are these:
\begin{equation}
c^{+}_{\lambda,m}
=\gamma_{m-\mu}\, \sqrt{\frac{\Gamma(\lambda+m)}{\Gamma(1-\lambda+m)}},\qquad
c^{-}_{\lambda,m}
=\gamma_{m-\mu}\, \sqrt{\frac{\Gamma(\lambda-m)}{\Gamma(1-\lambda-m)}}\qquad
\text{\begin{tabular}{c}(using positive\\ square roots).\end{tabular}}
\end{equation}
For the representation $\calD_\lambda^+$ or $\calD_\lambda^-$, one of the above equations is applicable. More explicitly,
\begin{equation}
c^{+}_{\lambda,\,\lambda+k}
=\sqrt{\frac{\Gamma(2\lambda+k)}{k!}}
\quad\text{ or }\quad
c^{-}_{\lambda,\,-(\lambda+k)}
=(-1)^k \sqrt{\frac{\Gamma(2\lambda+k)}{k!}}\qquad\:
(k=0,1,2,\ldots).
\end{equation}
For other values of $m$, the numbers $c^{\pm}_{\lambda,m}$ vanish.

\section{Fourier transform on $\tGG$ and the Plancherel measure}\label{sec_Fourier}

We first remind the reader of some algebraic terminology. An \emph{action} of a group $G$ on a set $X$ is a function $A:\,G\times X\to X$ such that $A(1,x)=x$ and $A(gh\,,x)=A\bigl(g,\,A(h,x)\bigr)$ for all $g$, $h$, and $x$. Instead of $A(g,x)$, it is more common to write $A(g)\, x$ or $A(g)\cdot x$. This construction is also called a ``left group action''. A ``right'' action is a similar function $B$ that satisfies the equation $B(hg,\,x)=B(g,\,B(h,x))$. Any right action $B$ can be turned into the left action $A(g,x)=B(g^{-1},x)$. Thus, the concept of a right action is redundant when working with groups (as opposed to semigroups).

Invertible maps $V:\,X\to X$ act on functions as follows:
\begin{equation}
(Vf)(x)=f(V^{-1}(x)).
\end{equation}
Any group $G$ acts on itself in two different ways:
\begin{equation}
\La(h)\cdot g=hg,\qquad \Ra(h)\cdot g=gh^{-1}.
\end{equation}
We call the (left) actions $\La$, $\Ra$ the \emph{L-action} and \emph{R-action}, respectively. They can also be applied to functions on $G$. If $f$ is such a function, then
\begin{equation}
\bigl(\La(h)\cdot f\bigr)(g)=f(h^{-1}g),\qquad
\bigl(\Ra(h)\cdot f\bigr)(g)=f(gh).
\end{equation}

\subsection{Fourier transform on a group}\label{sec_Fourier1}

Let $G$ be a group. We denote its unitary irreps by $\alpha,\beta$, etc., the corresponding Hilbert spaces by $\calU_{\alpha}$, and the group actions on them by $U_\alpha$. Given some orthonormal basis of $\calU_{\alpha}$, the \emph{matrix element functions} $\Up{\alpha}{j}{k}$ and $\Um{\alpha}{j}{k}$ are defined as follows:
\begin{equation}
\wideboxed{
\Up{\alpha}{j}{k}(g)=\bbra{j}U_{\alpha}(g)\bket{k},\qquad\quad
\Um{\alpha}{j}{k}(g)=\Up{\alpha}{j}{k}(g^{-1})
=\bbra{k}U_{\alpha}(g)\bket{j}^{*}
}
\end{equation}
The latter are slightly more convenient as a basis of the space of functions on $G$. Let us first write some elementary things, namely, the composition law and the group action on ket- and bra-vectors in the matrix notation:
\useshortskip{\begin{gather}
\Up{\alpha}{j}{k}(gh)
=\sum_{n}\,\Up{\alpha}{j}{n}(g)\:\Up{\alpha}{n}{k}(h),
\displaybreak[0]\\[3pt]
U_{\alpha}(h)\,\ket{j}=\sum_{n}\,\Up{\alpha}{n}{j}(h)\,\ket{n},
\qquad\quad
\bra{k}\,U_{\alpha}(h^{-1})=\sum_{n}\,\Um{\alpha}{k}{n}(h)\,\bra{n}.
\end{gather}}
(In the case of bra-vectors, we use $h^{-1}$ to conform to the definition of the group action.) Now, one can easily check that the functions $\Um{\alpha}{k}{j}$ are transformed under the L- and R-actions as $\ket{j}\bra{k}\in\calU_\alpha\otimes\calU_\alpha^*$:
\begin{equation}
\La(h)\cdot\Um{\alpha}{k}{j}
=\sum_{n}\,\Up{\alpha}{n}{j}(h)\: \Um{\alpha}{k}{n},\qquad\quad
\Ra(h)\cdot\Um{\alpha}{k}{j}
=\sum_{n}\,\Um{\alpha}{k}{n}(h)\: \Um{\alpha}{n}{j}.
\end{equation}

If $G$ is compact, the matrix element functions $\Um{\alpha}{k}{j}$ satisfy the Schur orthogonality relation, where the inner product is defined by the (arbitrarily normalized) Haar measure. Let $d_\alpha$ be the dimension of $\alpha$; then the orthogonality relation is:
\begin{equation}\label{ortho_comp}
\bbraket{\Um{\alpha}{k}{j}}{\Um{\beta}{l}{m}}
=\int_{G}
\bbra{j}U_{\alpha}(g)\bket{k}\,\bbra{l}U_{\beta}(g^{-1})\bket{m}\, dg
=\bigl({\textstyle d_\alpha/\int_{G}dg}\bigr)^{-1}
\delta_{\alpha\beta}\,\delta_{kl}\,\delta_{jm}.
\end{equation}
Furthermore, the functions $\Um{\alpha}{k}{j}$ form a basis of the Hilbert space $\calH$ of functions on $G$. The number $d_\alpha/\int_{G}dg$ is called the \emph{Plancherel measure} of the irrep $\alpha$. It appears as a coefficient in the decomposition of identity into the projectors onto the basis functions. For example, if $G$ is finite, the integral becomes a sum, and the Plancherel measure is $d_\alpha/|G|$. The decomposition into matrix element functions is known as the \emph{Fourier transform on $G$}. 

The generalization of these results to non-compact Lie groups, such as $\tGG$, is not straightforward. To understand the problems that can and do arise, let us prove the orthogonality and completeness for finite groups. To show the orthogonality, let $M^{\alpha,jl}_{\beta,km}$ denote the middle expression in~(\ref{ortho_comp}). If $\alpha,\beta,k,l$ are fixed, then the expression
\begin{equation}
\sum_{j,m}M^{\alpha,jl}_{\beta,km}\ket{j}\bra{m}
=\int_{G} U_{\alpha}(g)\bket{k}\,\bbra{l}U_{\beta}(g^{-1})\, dg
\end{equation}
defines a linear map from $\calU_\beta$ to $\calU_\alpha$, which is easily seen to be an intertwiner. By Schur's lemma, it is proportional to the identity map if $\alpha=\beta$ and vanishes otherwise. Hence, $M^{\alpha,jl}_{\beta,km}$ is proportional to $\delta^{\alpha}_{\beta}\delta^{j}_{m}$ and, by a similar argument, to $\delta^{\alpha}_{\beta}\delta^{l}_{k}$. These two statements imply that $M^{\alpha,jl}_{\beta,km} \propto\delta^{\alpha}_{\beta}\delta^{j}_{m}\delta^{l}_{k}$. The proportionality coefficient can be found by contracting the indices $k$ and $l$.

To show the completeness, we consider the Hilbert space $\calH$ as the left regular representation of $G$ (that is, the representation given by the L-action) and decompose it into unitary irreps. Any of these sub-representations has a basis of functions $f_j$ that transform as the basis vectors $\ket{j}$ of some standard representation $\alpha$. Hence,
\begin{equation}
f_{j}(g)=\bigl(\La(g^{-1})\cdot f_j\bigr)(1)
=\sum_{n}\Um{\alpha}{n}{j}(g)\,f_{n}(1),
\end{equation}
which means that $f_j$ is the linear combination of the functions $\Um{\alpha}{n}{j}$ with the coefficients $f_{n}(1)$. Therefore, any $f\in\calH$ is a combination of matrix element functions.

What changes if $G$ is a general locally compact Lie group? The use of Schur's lemma is valid if the left and right Haar measures are the same (which is true for $G=\tGG$). However the integral in Eq.~(\ref{ortho_comp}) may diverge. For example, if $G=\RR$, then $\oU_\alpha(g)=e^{i\alpha g}$ (for $\alpha,g\in\RR$). In this case, the functions $\oU_\alpha$ are $\delta$-normalizable, namely $\braket{\oU_\alpha}{\oU_\beta}=2\pi\delta(\alpha-\beta)$. This gives the decomposition of identity $\mathbf{1}=(2\pi)^{-1}\int\ket{\oU_\alpha}\bra{\oU_\alpha}\,d\alpha$; thus, the Plancherel measure on the set of irreps is $(2\pi)^{-1}d\alpha$. The trivial representation of $\tGG$ is a different case. The corresponding function $\oU_{0}$ is identically equal to $1$, and its norm is clearly infinite. One might hope to remedy the situation by regarding the trivial representation as a limiting case of the complementary series $\calC^{0}_{q}$ for $q\to0$ or the discrete series $\calD_{\lambda}^{\pm}$ for $\lambda\to0$. However, it turns out that all complementary series representations and the discrete series representations with $\lambda<\frac{1}{2}$ have matrix elements that are not even $\delta$-normalizable. Therefore, these representations do not appear in the Fourier transform.

The main assumption in the completeness proof was the existence of an irreducible decomposition of the left regular representation. It is true that any unitary representation of $\tGG$ splits into irreducible pieces, which are isomorphic to the standard unitary irreps. However, this fact should not be taken for granted. For some other groups (e.g.\ $\SL(2,\ZZ)$), a general unitary representation does not split into irreducible representations (because the process of splitting into progressively smaller pieces may not converge). A more general decomposition into isotypical components exists, but it involves type II and type III von Neumann factors. We will not prove or use the existence of an irreducible decomposition for an arbitrary unitary representation of $\tGG$ but show the completeness of the matrix element functions directly.

\subsection{The L- and R-actions on $\tGG$ in explicit coordinates}\label{sec_coord_tGG}

It is convenient to parametrize $\tGG$ by three variables similar to the Euler angles:
\begin{equation}\label{gparam}
\wideboxed{
g(\xi,\ph,\vth)=e^{\ph\Lambda_0}e^{\xi\Lambda_1}e^{-\vth\Lambda_0},\qquad
\xi\ge 0
}
\end{equation}
Note that $g(\xi,\ph+2\pi,\vth+2\pi)=g(\xi,\ph,\vth)$. For a nonsingular, one-to-one parametrization, one can use $z=e^{i\ph}\tanh(\xi/2)$ (the image of $0$ under the action of $g(\xi,\ph,\vth)$ on the unit disk) and $\vth-\ph$.

Let us find the L- and R-actions in the infinitesimal form, i.e.\ calculate $(1+\delta h)g$ and $g(1-\delta h)$, where $\delta h$ is a Lie algebra element and $g\in\tGG$. When $\delta h$ is fixed, the expressions $(\delta h)g$ and $g(\delta h)$ define some vector fields on $\tGG$. In general,
\begin{equation}
(\delta h)g=X^\la(g)\,(\delta h),\qquad
g(\delta h)=X^\ra(g)\,(\delta h),
\end{equation}
where $X^\la(g)$ and $X^\ra(g)$ are linear maps from the Lie algebra to the tangent space of $\tGG$ at point $g$. They are represented by matrices if we write vector fields in components and express $\delta h$ as $(\delta h^j)\Lambda_j$. Thus, the L- and R-actions of $\Lambda_j$ on functions are given by these formulas:
\begin{equation}\label{LRfunc}
\Lambda_j^\la=-\bigl[X^\la(g)\bigr]_j^\alpha\,
\frac{\partial}{\partial g^\alpha},\qquad
\Lambda_j^\ra=\bigl[X^\ra(g)\bigr]_j^\alpha\,
\frac{\partial}{\partial g^\alpha},
\end{equation}
where $j=0,1,2$ and the index $\alpha$ refers to $\xi$, $\ph$, or $\vth$.

It is easier to calculate the inverse matrices, $Y^\la(g)=(X^\la(g))^{-1}$ and $Y^\ra(g)=(X^\ra(g))^{-1}$, which represent two variants of the Maurer-Cartan form:
\begin{equation}
(dg)g^{-1}=\Lambda_{j}\,\bigl[Y^\la(g)\bigr]^{j}_{\alpha}\,dg^\alpha,\qquad
g^{-1}(dg)=\Lambda_{j}\,\bigl[Y^\ra(g)\bigr]^{j}_{\alpha}\,dg^\alpha.
\end{equation}
To obtain, for example, the first column of $Y^\la=Y^\la(g)$ for $g=g(\xi,\ph,\vth)$, we consider
\[
\frac{\partial g}{\partial\xi}\,g^{-1}
=\bigl(e^{\ph\Lambda_0}\Lambda_{1}e^{\xi\Lambda_1}e^{-\vth\Lambda_0}\bigr)
\bigl(e^{\ph\Lambda_0}e^{\xi\Lambda_1}e^{-\vth\Lambda_0}\bigr)^{-1}
=e^{\ph\Lambda_0}\Lambda_{1}e^{-\ph\Lambda_0}
=(\cos\ph)\Lambda_{1}+(\sin\ph)\Lambda_{2},
\]
and collect the coefficients in front of $\Lambda_j$. Thus, the first column of $Y^\la$ is $(0,\cos\ph,\,\sin\ph)^T$. Continuing in this manner, we find that
\begin{align}
Y^\la &=\begin{pmatrix}
0 & 1 & -\cosh\xi\\
\cos\ph & 0 & -\sin\ph\, \sinh\xi\\
\sin\ph & 0 & \cos\ph\, \sinh\xi
\end{pmatrix}, &
Y^\ra &=\begin{pmatrix}
0 & \cosh\xi & -1\\
\cos\vth & -\sin\vth\, \sinh\xi & 0\\
\sin\vth & \cos\vth\, \sinh\xi & 0
\end{pmatrix},
\displaybreak[0]\\[10pt]
\label{XLR}
X^\la &=\begin{pmatrix}
0 & \cos\ph & \sin\ph
\\[2pt]
1 & -\sin\ph\,\dfrac{\cosh\xi}{\sinh\xi} & \cos\ph\,\dfrac{\cosh\xi}{\sinh\xi}
\vspace{3pt}\\
0 & -\sin\ph\,\dfrac{1}{\sinh\xi} & \cos\ph\,\dfrac{1}{\sinh\xi}
\end{pmatrix}, &
X^\ra &=\begin{pmatrix}
0 & \cos\vth & \sin\vth
\\[2pt]
0 & -\sin\vth\,\dfrac{1}{\sinh\xi} & \cos\vth\,\dfrac{1}{\sinh\xi}
\vspace{3pt}\\
-1 & -\sin\vth\,\dfrac{\cosh\xi}{\sinh\xi} & \cos\vth\,\dfrac{\cosh\xi}{\sinh\xi}
\end{pmatrix}.
\end{align}\enlargethispage{10pt}

We can now plug the expressions for $X^\la$, $X^\ra$ into Eq.~(\ref{LRfunc}) to obtain the L- and R-actions of the operators $\Lambda_j$ on functions. It is convenient to have the result for $L_{0}=-i\Lambda_0$ and $L_{\pm1}=\mp\Lambda_1-i\Lambda_2$:
\begin{empheq}[box=\widebox]{alignat=2}
\label{Laction}
L_{0}^\la &=i\partial_\ph,\qquad &
L_{\pm1}^\la &=e^{\pm i\ph}
\left( \pm\partial_\xi
+\frac{\cosh\xi}{\sinh\xi}\,(i\partial_\ph)
+\frac{1}{\sinh\xi}\,(i\partial_\vth) \right)
\\[5pt]
\label{Raction}
L_{0}^\ra &=i\partial_\vth,\qquad &
L_{\pm1}^\ra &=e^{\pm i\vth}
\left( \mp\partial_\xi
-\frac{1}{\sinh\xi}\,(i\partial_\ph)
-\frac{\cosh\xi}{\sinh\xi}\,(i\partial_\vth) \right)
\end{empheq}

\subsection{Casimir eigenfunctions}\label{sec_Casimir_ef}

The space of square-integrable functions on $\tGG$ can be decomposed into common eigenfunctions of three commuting Hermitian operators, $L_{0}^\la$, $L_{0}^\ra$, and $Q$, where
\begin{equation}
Q=-(L_0^\la)^2
+\frac{1}{2}\bigr(L_{-1}^\la L_{1}^\la+L_{1}^\la L_{-1}^\la\bigr)
=-(L_0^\ra)^2
+\frac{1}{2}\bigr(L_{-1}^\ra L_{1}^\ra+L_{1}^\ra L_{-1}^\ra\bigr).
\end{equation}
We now find the common eigenfunctions without asking if they are normalizable. That question will be addressed later.

Let us first impose the conditions $L_{0}^\la\Psi=-l\Psi$ and $L_{0}^\ra\Psi=-r\Psi$, where $l$ and $r$ are arbitrary complex numbers.\footnote{The parameter $\nu=-r$ may be called ``spin'' because a function $\Psi$ satisfying the condition $L_{0}^\ra\Psi=\nu\Psi$ has the interpretation as a $\nu$-spinor on the hyperbolic plane, see section~\ref{sec_spinors}.} Thus,
\begin{equation}\label{Psi_f}
\Psi\bigl(e^{\ph\Lambda_0}e^{\xi\Lambda_1}e^{-\vth\Lambda_0}\bigr)
=e^{i(l\ph+r\vth)}f(u),\qquad \text{where}\quad u=\tanh^2\frac{\xi}{2}.
\end{equation}
The use of the variable $u$ instead of $\xi$ will help to simplify some subsequent equations. In what follows, $\Psi$ is treated as a function of $(\xi,\ph,\vth)$ rather than $g=e^{\ph\Lambda_0}e^{\xi\Lambda_1}e^{-\vth\Lambda_0}$. We will later require that $\Psi(g)$ only depend on $g$ and be regular at $g=1$.

The action of $L_n^\la$, $L_n^\ra$ on $f$ depends on the parameters $l$ and $r$,
\begin{gather}
L_{0}^\la(l,r) =-l, \qquad\quad
L_{\pm1}^\la(l,r) =\pm(1-u)u^{1/2}\partial_{u}
-\frac{l+r}{2}\,u^{-1/2}-\frac{l-r}{2}\,u^{1/2},
\\[5pt]
\label{LRlr2}
L_{n}^\ra(l,r) = (-1)^{n}L_{n}^\la(r,l),
\end{gather}
and the parameters also change:
\begin{equation}
L_{n}^\la:\,(l,r)\mapsto(l+n,r),\qquad
L_{n}^\ra:\,(l,r)\mapsto(l,r+n).
\end{equation}
It is now easy to write the Casimir operator explicitly:
\begin{equation}
Q=-(1-u)^2\,\bigl(u\partial_{u}^2+\partial_{u}\bigr)
+\frac{1-u}{4u}\Bigl((l+r)^2-(l-r)^{2}u\Bigr).
\end{equation}
The eigenvalue equation, $Qf=\lambda(1-\lambda)f$ is equivalent to the hypergeometric differential equation $(u\partial_u+c)\partial_uh =(u\partial_u+a)(u\partial_u+b)h$ for a closely related function $h$. Indeed, both differential equations have regular singular points at $u=0,1,\infty$ and no other singularities. To find the exact relation, it is sufficient to compare the characteristic exponents that define the asymptotics of the fundamental solutions, namely, $f(u)\sim u^{\alpha_v}$ for $u\to v$\, ($v=0,1$) and $f(u)\sim u^{-\alpha_\infty}$ for $u\to\infty$. The Casimir eigenvalue equation and the hypergeometric equation have the following exponents:
\begin{equation}
\begin{array}{@{}l}
\text{Equation for $f$}\\
\text{with parameters}\\
\lambda,\,l,\,r\,:
\end{array}
\left\{\begin{array}{@{}l@{}}
\alpha_0=\pm(l+r)/2,\\
\alpha_1=\lambda,\,1-\lambda,\\
\alpha_\infty=\pm(l-r)/2;
\end{array}\right.\qquad\qquad
\begin{array}{@{}l}
\text{Equation for $h$}\\
\text{with parameters}\\
 a,\,b,\,c\,:
\end{array}
\left\{\begin{array}{@{}l@{}}
\alpha_0=0,\,1-c,\\
\alpha_1=0,\,c-a-b,\\
\alpha_\infty=a,\,b.
\end{array}\right.
\end{equation}
Since each exponent has two different values, there are several ways to match them. For example,
\begin{equation}
f(u)=u^{(l+r)/2}(1-u)^{\lambda}h(u),\qquad
a=\lambda+l,\quad b=\lambda+r,\quad c=1+l+r.
\end{equation}

\paragraph{General solutions:} These are two solutions of the equation $Qf=\lambda(1-\lambda)f$ on the interval $0<u<1$:
\begin{empheq}[box=\widebox]{align}
\label{Adef}
A_{\lambda,l,r}(u)&=u^{(l+r)/2}(1-u)^{\lambda}\,
\hgfs\bigl(\lambda+l,\,\lambda+r,\,1+l+r;\,u\bigr)\\[5pt]
\label{Bdef}
B_{\lambda,l,r}(u)&=u^{(l+r)/2}(1-u)^{\lambda}\,
\hgfs\bigl(\lambda+l,\,\lambda+r,\,2\lambda;\,1-u\bigr)
\end{empheq}
where $\hgfs(a,b,c;x)=\Gamma(c)^{-1}\hgf(a,b,c;x)$ is the scaled hypergeometric function. (It is well-defined for all values of $a$, $b$, $c$ but vanishes if $a,c-a\in\{0,-1,-2,\ldots\}$ or $b,c-b\in\{0,-1,-2,\ldots\}$.) Let us mention some useful identities:
\begin{gather}
\label{ABsymm}
A_{1-\lambda,l,r}=A_{\lambda,l,r},\qquad
B_{\lambda,l,r}=B_{\lambda,-l,-r};\\[6pt]
\label{ABconn}
\frac{\sin(2\pi\lambda)}{\pi}\, A_{\lambda,l,r}
=\frac{B_{\lambda,l,r}}{\Gamma(1-\lambda+l)\,\Gamma(1-\lambda+r)} 
-\frac{B_{1-\lambda,l,r}}{\Gamma(\lambda+l)\,\Gamma(\lambda+r)}\, .
\end{gather}
For a more complete picture, $A_{\lambda,l,r}(u)$, $A_{\lambda,-l,-r}(u)$ make a pair of fundamental solutions near $u=0$, the functions $B_{\lambda,l,r}(u)$, $B_{1-\lambda,l,r}(u)$ are the fundamental solutions near $u=1$, and $A_{\lambda,l,-r}(u^{-1})$, $A_{\lambda,-l,r}(u^{-1})$ near $u=\infty$. (The first four functions are defined for $u\in(0,1)$ and the last two for $u\in(1,\infty)$.) The operators $L_{\pm1}^\la(l,r)$ act on $A_{\lambda,l,r}$ and $B_{\lambda,l,r}$ as follows:
\begin{align}
\label{La_act_A}
L_{-1}^\la(l,r)\, A_{\lambda,l,r}&=-A_{\lambda,l-1,r}\,, &
L_{1}^\la(l,r)\, A_{\lambda,l,r}
&=-(l+\lambda)(l+1-\lambda)A_{\lambda,l+1,r}\,;\\[5pt]
\label{La_act_B}
L_{-1}^\la(l,r)\, B_{\lambda,l,r}&=-(l-\lambda)B_{\lambda,l,r}\,, &
L_{1}^\la(l,r)\, B_{\lambda,l,r}&=-(l+\lambda)B_{\lambda,l,r}\,.
\end{align}
To find the action on the other fundamental solutions, we note that $L_{n}^\la(l,r)=-L_{-n}^\la(-l,-r)$ and that $L_{n}^\la(l,r)$ acts on functions of $u^{-1}$ as $L_{n}^\la(l,-r)$ on functions of $u$. The R-action is obtained from Eq.~(\ref{LRlr2}).

In some applications (e.g.\ spinors on $\widetilde{\AdS}_2$, where $u=e^{i(\ph_1-\ph_2)}$), $u$ lies on the unit circle. One can analytically continue functions from the interval $(0,1)$ to the simply connected domain $D=\CC-[0,\infty)$ (containing the unit circle without the point $1$) through the upper half-plane or through the lower half-plane. The first option is preferred when the circle is parametrized as $u=e^{i\theta}$ with $0<\theta<2\pi$. However, let us give both definitions and some related identities:
\begin{gather}
\left.\begin{array}{@{}c@{}}
A_{\lambda,l,r}^{\pm} \\[3pt] B_{\lambda,l,r}^{\pm}
\end{array}\right\}
=\text{analytic cont.\ of}
\left\{\begin{array}{@{}c@{}}
A_{\lambda,l,r} \\[3pt] B_{\lambda,l,r}
\end{array}\right\}
\text{through the upper ($+$) or lower ($-$) half-plane};
\displaybreak[0]\\[8pt]
e^{-i\pi\frac{l+r}{2}}A_{\lambda,l,r}^{+}(u)
=e^{i\pi\frac{l+r}{2}}A_{\lambda,l,r}^{-}(u)
,\hspace{1.5cm}\\[6pt]
e^{i\frac{\pi}{2}\lambda} B_{\lambda,l,r}^{+}(u)
=e^{i\frac{\pi}{2}\lambda} B_{\lambda,-l,-r}^{+}(u)
=e^{-i\frac{\pi}{2}\lambda} B_{\lambda,l,-r}^{-}(u^{-1})
=e^{-i\frac{\pi}{2}\lambda} B_{\lambda,-l,r}^{-}(u^{-1}).
\end{gather}
We now write all 6 fundamental solutions that are continued from their original definition domains to $D$ through the half-plane $\Im u>0$:
\begin{equation}
\begin{aligned}
&A_{\lambda,l,r}^{+}(u), & &B_{\lambda,l,r}^{+}(u), &
&A_{\lambda,l,-r}^{-}(u^{-1}),\\[3pt]
&A_{\lambda,-l,-r}^{+}(u), & &B_{1-\lambda,l,r}^{+}(u), &
&A_{\lambda,-l,r}^{-}(u^{-1}).
\end{aligned}
\end{equation}
They span the two-dimensional solution space.

\paragraph{Nonsingular solutions:} Having studied the Casimir eigenfunctions of the form $e^{i(l\ph+r\vth)}f(u)$ in full generality, we select those that depend only on $g=g(\xi,\ph,\vth)\in\tGG$ and are regular at $g=1$. The first condition means the invariance under $(\xi,\ph,\vth)\mapsto(\xi,\ph+2\pi,\vth+2\pi)$; hence, $l+r$ is an integer. To check if a function is regular, we examine its $u\to 0$ asymptotics using the variables $z=e^{i\ph}u^{1/2}$,\, $\bar{z}=e^{-i\ph}u^{1/2}$, and $\vth-\ph$. The space of regular solutions is spanned by $A_{\lambda,l,r}$ and $A_{\lambda,-l,-r}$. These functions are linearly dependent because
\begin{equation}\label{Alindep}
\Gamma(\lambda+l)\,\Gamma(\lambda+r)\,A_{\lambda,l,r}
=\Gamma(\lambda-l)\,\Gamma(\lambda-r)\,A_{\lambda,-l,-r}\qquad
\text{if }\, l+r\in\ZZ.
\end{equation}
One of them may vanish, but $A_{\lambda,l,r}\not=0$ if $l+r\ge 0$ and $A_{\lambda,-l,-r}\not=0$ if $l+r\le 0$.

\paragraph{Normalizability:} Finally, we select the nonsingular solutions that are normalizable or $\delta$-nor\-ma\-lizable. The inner product is given by the Haar measure on $\tGG$. When group elements are represented as $e^{i\ph\Lambda_0}e^{i\xi\Lambda_1}e^{-i\vth\Lambda_0}$, the measure is $(\sinh\xi)\,d\xi\,d\ph\,d\vth$. If $\Psi_{\alpha}(g) =e^{i(l_\alpha\ph+r_\alpha\vth)} f_{\alpha}\bigl(\tanh^2(\xi/2)\bigr)$ with $l_\alpha+r_\alpha\in\ZZ$ and $l_\alpha,r_\alpha\in\RR$ (where $\alpha=1,2$), then
\begin{equation}\label{ip_tGG}
\braket{\Psi_1}{\Psi_2}=4\pi^2\delta_{l_1+r_1,\,l_2+r_2}\delta(r_1-r_2)\,
\braket{f_1}{f_2},\quad\: \text{where}\quad
\braket{f_1}{f_2}=\int_{0}^{1} f_1(u)^{*}f_2(u)\,\frac{2\,du}{(1-u)^2}.
\end{equation}

Since $\lambda$ and $1-\lambda$ define the same Casimir eigenspace, we may assume that $\Re\lambda>\frac{1}{2}$, or $\lambda=\frac{1}{2}+is$ with $s>0$, or $\lambda=\frac{1}{2}$. Let us consider one of the linearly dependent candidate solutions $A_{\lambda,l,r}$ and $A_{\lambda,-l,-r}$:
\begin{equation}
\label{Aasympt}
A_{\lambda,l,r}(u)
\approx a_{\lambda,l,r}(1-u)^{\lambda}+a_{1-\lambda,l,r}(1-u)^{1-\lambda}
\quad\:\text{for }\, u\to 1,
\end{equation}
where
\begin{equation}\label{a_llr}
a_{\lambda,l,r}
=\frac{\Gamma(1-2\lambda)}{\Gamma(1-\lambda+l)\,\Gamma(1-\lambda+r)}.
\end{equation}
If $\Re\lambda>\frac{1}{2}$, a function with the asymptotic behavior $f(u)\sim(1-u)^{\lambda}$ for $u\to 1$ is normalizable, and $(1-u)^{1-\lambda}$ is not even $\delta$-normalizable. In the marginal case of $\lambda=\frac{1}{2}+is$,\,\,$s>0$, we have:
\begin{equation}
\text{if }\, f_{s}(u) \approx a(1-u)^{1/2+is}+a^{*}(1-u)^{1/2-is}\,
\text{ for }\, u\to 1,\qquad
\text{then }\, \braket{f_{s}}{f_{s'}}=4\pi|a|^2\delta(s-s').
\end{equation}
If $\lambda=\frac{1}{2}$, one has to take the limit in Eq.~(\ref{Aasympt}); the result is that the function is not normalizable. It is now easy to find all cases where both $A_{\lambda,l,r}$ and $A_{\lambda,-l,-r}$ are normalizable or $\delta$-normalizable:
\begin{align}
\label{norm_ef}
\bullet&\quad\: \lambda>\frac{1}{2}\quad \text{and}\quad
\Bigl(\lambda+l,\,\lambda-r\in\{0,-1,-2,\ldots\}\, \text{ or }\,
\lambda+r,\,\lambda-l\in\{0,-1,-2,\ldots\}\Bigr);\\[3pt]
\label{dnorm_ef}
\bullet&\quad\: \lambda=\frac{1}{2}+is,\quad\:
s,l,r\in\RR,\quad\: s>0,\quad\: l+r\in\ZZ.
\end{align}

\subsection{The matrix elements and Plancherel measure for the irreps of $\tGG$}

Let $\calU^{\mu}_{\lambda}$ be a nontrivial unitary irrep, i.e.\ $\calC^{\mu}_{\lambda(1-\lambda)}$ or $\calD^{\pm}_{\lambda}$. The corresponding matrix element functions $\Um{\lambda}{\nu}{m}$ (for $m,\nu\in\ZZ+\mu$) transform as $\ket{m}\bra{\nu}$ under the L- and R-actions of the group. In particular,
\begin{gather}
L_0^{\la}\,\Um{\lambda}{\nu}{m}=-m\Um{\lambda}{\nu}{m},\qquad
L_0^{\ra}\,\Um{\lambda}{\nu}{m}=\nu\Um{\lambda}{\nu}{m},\qquad
Q\,\Um{\lambda}{\nu}{m}=\lambda(1-\lambda)\Um{\lambda}{\nu}{m},
\\[5pt]
L_{\pm1}^{\la}\,\Um{\lambda}{\nu}{m}
=-\sqrt{(m\pm\lambda)(m\pm(1-\lambda))}\,\Um{\lambda}{\nu}{m\pm1}.
\end{gather}
The first set of equations implies that $\Um{\lambda}{\nu}{m} \bigl(e^{\ph\Lambda_0}e^{\xi\Lambda_1}e^{-\vth\Lambda_0}\bigr) =e^{i(m\ph-\nu\vth)}f\bigl(\tanh^2(\xi/2)\bigr)$, where $f$ is proportional to the fundamental solution $A_{\lambda,m,-\nu}$ or $A_{\lambda,-m,\nu}$. Comparing the action of $L_{\pm1}^{\la}$ on the functions $\Um{\lambda}{\nu}{m}$ with the corresponding action~(\ref{La_act_A}) on the fundamental solutions and using the identity $\Um{\lambda}{\nu}{m}(1)=\delta^{\nu}_{m}$ for normalization, we find that
\begin{equation}\label{U_lnm}
\wideboxed{
\begin{aligned}
\Um{\lambda}{\nu}{m} \Bigl(e^{\ph\Lambda_0}e^{\xi\Lambda_1}e^{-\vth\Lambda_0}\Bigr)
={}&\sqrt{\frac{\Gamma(\lambda+m)\,\Gamma(1-\lambda+m)}
{\Gamma(\lambda+\nu)\,\Gamma(1-\lambda+\nu)}}\,
e^{i(m\ph-\nu\vth)} A_{\lambda,m,-\nu}\biggl(\tanh^2\frac{\xi}{2}\biggr)
\\[3pt]
{}=(-1)^{\nu-m}&\sqrt{\frac{\Gamma(\lambda-m)\,\Gamma(1-\lambda-m)}
{\Gamma(\lambda-\nu)\,\Gamma(1-\lambda-\nu)}}\,
e^{i(m\ph-\nu\vth)} A_{\lambda,-m,\nu}\biggl(\tanh^2\frac{\xi}{2}\biggr)
\end{aligned}
}
\end{equation}
The first formula is applicable to the irreps $\calC^{\mu}_{\lambda(1-\lambda)}$, $\calD^{+}_{\lambda}$, and the second to $\calC^{\mu}_{\lambda(1-\lambda)}$, $\calD^{-}_{\lambda}$. Note that the range of $m$ and $\nu$ is restricted to $\{\lambda,\lambda+1,\ldots\}$ for $\calD_\lambda^+$ and to $\{-\lambda,-\lambda-1,\ldots\}$ for $\calD_\lambda^-$. Thus, the normalizable and $\delta$-normalizable Casimir eigenfunctions (see equations ~(\ref{norm_ef}), (\ref{dnorm_ef})) are exactly the matrix element functions for $\calD_\lambda^\pm$ with $\lambda>\frac{1}{2}$ and for $\calC^{\mu}_{\lambda(1-\lambda)}$ with $\lambda=\frac{1}{2}+is$. This shows the completeness of the matrix element functions.

The analogue of orthogonality relation~(\ref{ortho_comp}) for discrete series representations is as follows:
\begin{equation}\label{ortho_discr}
\begin{gathered}
\Bigl\langle\Um{\lambda}{\pm(\lambda+k)}{\,\pm(\lambda+j)}\,\Big|\,
\Um{\lambda'}{\pm(\lambda'+k')}{\,\pm(\lambda'+j')}\Bigr\rangle
=\frac{8\pi^2}{2\lambda-1}\,\delta(\lambda-\lambda')\,
\delta_{jj'}\,\delta_{kk'}
\\[3pt]
\text{for}\quad \lambda,\lambda'>\frac{1}{2},\quad\:
j,k,j',k'\in\{0,1,2,\ldots\}.
\end{gathered}
\end{equation}
The overall factor is found by setting $j,k,j',k'$ to $0$ so that the functions in questions are $e^{\pm i\lambda(\ph-\vth)}(1-u)^{\lambda}$ and $e^{\pm i\lambda'(\ph-\vth)}(1-u)^{\lambda'}$, where $u=\tanh^2(\xi/2)$. The inner product between these functions is obtained using Eq.~(\ref{ip_tGG}). For the principal series, the orthogonality relation is:
\begin{equation}\label{ortho_cont}
\begin{gathered}
\Bigl\langle\Um{1/2+is}{\mu+k}{\,\mu+j}\,\Big|\,
\Um{1/2+is'}{\mu'+k'}{\,\mu'+j'}\Bigr\rangle
=4\pi^2\,\frac{\cosh(2\pi s)+\cos(2\pi\mu)}{s\sinh(2\pi s)}\,
\delta(s-s')\,\delta(\mu-\mu')\,\delta_{jj'}\,\delta_{kk'}
\\[3pt]
\text{for}\quad s,s'>0,\quad\:
-\frac{1}{2}<\mu,\mu'\le\frac{1}{2},\quad\: j,k,j',k'\in\ZZ.
\end{gathered}
\end{equation}
(The range of $s$ and $\mu$ has been restricted to avoid redundancy.)
To derive equation~(\ref{ortho_cont}), we again consider the case $j=k=j'=k'=0$. The function $\Um{1/2+is}{\mu}{\,\mu}$ is proportional to $A_{1/2+is,\,\mu,-\mu}$, which has the asymptotics~(\ref{Aasympt}) with
\begin{equation}
\bigl|a_{1/2+is,\,\mu,-\mu}\bigr|^2=\bigl|a_{1/2-is,\,\mu,-\mu}\bigr|^2
=\frac{\cosh(2\pi s)+\cos(2\pi\mu)}{4\pi s\sinh(2\pi s)}.
\end{equation}

The Plancherel measure is found by inverting the coefficients in the orthogonality relations:
\begin{equation}\label{Plancherel}
\wideboxed{
\begin{array}{@{}c@{}}
\text{Plancherel measure on}\\ \text{the irreps of $\tGG$}
\end{array}
\,=\begin{cases}
(2\pi)^{-2}\bigl(\lambda-\frac{1}{2}\bigr)\,d\lambda,\quad\:
\lambda>\tfrac{1}{2}
&\text{for }\, \calD_{\lambda}^{\pm}
\\[5pt]
(2\pi)^{-2}\dfrac{s\,\sinh(2\pi s)}{\cosh(2\pi s)+\cos(2\pi\mu)}
\,ds\,d\mu,\quad s>0
&\text{for }\, \calC_{1/4+s^2}^{\mu}
\end{cases}
}
\end{equation}
These formulas can be specialized to the irreps of $\GG\cong\PSL(2,\RR)$, which are characterized by $\mu=0$. In particular, the discrete series representations are $\calD_{n}^{\pm}$ with $n=1,2,\ldots$ Thus,
\begin{equation}
\wideboxed{
\begin{array}{@{}c@{}}
\text{Plancherel measure on}\\ \text{the irreps of $\PSL(2,\RR)$}
\end{array}
\,=\begin{cases}
(2\pi)^{-2}\bigl(n-\frac{1}{2}\bigr)
&\text{for }\, \calD_{n}^{\pm}
\\[3pt]
(2\pi)^{-2}\tanh(\pi s)\,s\,ds
&\text{for }\, \calC_{1/4+s^2}^{0}
\end{cases}
}
\end{equation}

\section{Spinors on the hyperbolic plane and anti-de Sitter  space}

\subsection{The spaces $\HH^2$, $\AdS_2$, and their complex embeddings}

The hyperbolic plane $\HH^2$ is the quotient of $\GG$ by the subgroup $K$ generated by $\Lambda_0$; it is also equal to the quotient of the corresponding universal covers, $\tGG/\tK$. Conversely, $\tGG$ is the total space of a principal $\tK$-bundle over $\HH^2$. Among the three ``Euler angle'' coordinates, $\xi$ and $\ph$ parametrize the base and $\vth$ the fiber. The metric on $\HH^2$ can be obtained from the L- and R-invariant metric on $\tGG$, which is in turn determined by the Killing form $\eta=\diag(-1,1,1)$. Using the notation of Section~\ref{sec_coord_tGG},
\begin{equation}
\begin{aligned}
d\ell^2&=\eta_{jk}(Y^\la)^j_\alpha(Y^\la)^k_\beta\:dg^\alpha\,dg^\beta
=\eta_{jk}(Y^\ra)^j_\alpha(Y^\ra)^k_\beta\:dg^\alpha\,dg^\beta\\[2pt]
&=d\xi^2-d\ph^2-d\vth^2+2\cosh\xi\,d\ph\,d\vth.
\end{aligned}
\end{equation}
The distance between infinitesimally close fibers is found by taking the extremum of $d\ell$ over $d\vth$ with $d\xi$ and $d\ph$ fixed. The result is:
\begin{equation}\label{Poincare_disk}
d\ell^2=d\xi^2+(\sinh\xi)^2 d\ph^2
=\frac{4\,dz\,d\bar{z}}{(1-z\bar{z})^2},\qquad\quad
\text{where}\quad
z=e^{i\ph}\tanh\frac{\xi}{2}.
\end{equation}
Thus, we have recovered the well-known Poincar\'e disk model on the hyperbolic plane. The space $\HH^2=\tGG/\tK$ inherits the L-action of $\tGG$, while the R-action has been used up in the quotient construction.

The anti-de Sitter space $\AdS_2$ is the quotient of $\GG$ by the subgroup generated by $\Lambda_2$. Recall that a general element of $\GG$ is a linear fractional map $g:\,z\mapsto\frac{az+b}{cz+d}$ preserving the unit disk. The subgroup generated by $\Lambda_2$ consists of those $g$'s that preserve $i$ and $-i$. Thus, $\AdS_2$ is the orbit of $(i,-i)$ under the simultaneous action of $\GG$ on pairs of points. This orbit, actually, includes all pairs of distinct points on the unit circle, $z_1=e^{i\ph_1}$ and $z_2=e^{i\ph_2}$. The standard projection $\GG\to\AdS_2$ takes $g$ to $(z_1,z_2)=(g(i),g(-i))$.

\begin{figure}
\centering\(\displaystyle
\figbox{1.0}{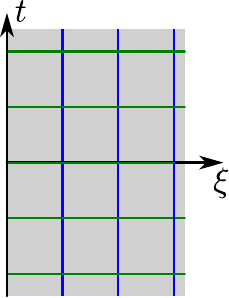}\quad\:
\quad\:\xrightarrow{\hspace{1.5cm}}\quad\:
\figbox{1.0}{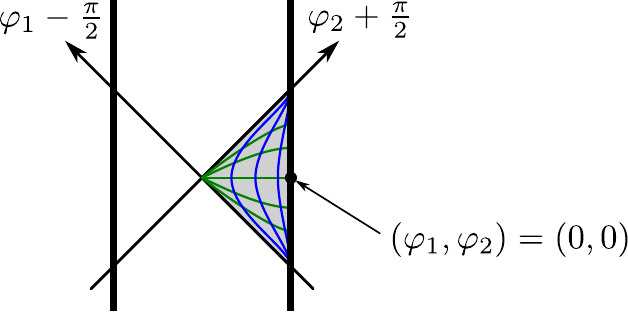}
\)
\caption{Schwarzschild patch of the anti-de Sitter space.}
\label{fig_AdS}
\end{figure}

To describe the metric on $\AdS_2$ and its universal cover $\widetilde{\AdS}_2$, we consider $\ph_1$, $\ph_2$ as real numbers subject to the constraint $0<\ph_1-\ph_2<2\pi$. Then
\begin{equation}
d\ell^2=d\xi^2-(\sinh\xi)^2 dt^2
=\frac{-\,d\ph_1\,d\ph_2}{\sin^2\bigl(\frac{\ph_1-\ph_2}{2}\bigr)},
\qquad\quad
\begin{aligned}
e^{-t}\tanh(\xi/2)&=\tan(\pi/4-\ph_1/2),\\[2pt]
e^{t}\tanh(\xi/2)&=\tan(\pi/4+\ph_2/2).
\end{aligned}
\end{equation}
(The first expression gives the metric in the region $\ph_1<\frac{\pi}{2}$,\,\, $\ph_2>-\frac{\pi}{2}$, which we call the ``Schwarzschild patch'', see Figure~\ref{fig_AdS}.) Another, more standard way to write the $\widetilde{\AdS}_2$ metric is $d\ell^2 =(\cos\theta)^{-2}(-d\ph^2+d\theta^2)$, where $\ph=(\ph_1+\ph_2)/2$ and $\theta=(\pi-\ph_1+\ph_2)/2$.

It is often useful to analytically continue functions between the hyperbolic plane and the anti-de Sitter space. From the physical perspective, there are two ways to define the analytic continuation: one is natural for the study of the SYK model in imaginary time (by interpreting a pair of points on the time circle as a point of $\AdS_2$~\cite{JSY16}) and the other corresponds to the transition to real time. We will obtain those continuations using a standard embedding of $\HH^2$ and two different embeddings of $\AdS_2$ in some complex manifold $\calM$. The latter consists of pairs of distinct points on the Riemann sphere $\overline{\CC}=\CC\cup\{\infty\}$ and comes with a complex metric:
\begin{equation}\label{spaceM}
\calM=\bigl\{(z_1,z_2):\: z_1,z_2\in\overline{\CC},\,\:z_1\not=z_2 \bigr\},
\qquad\quad
d\ell^2=\frac{-4\,dz_1\,dz_2}{(z_1-z_2)^2}.
\end{equation}
This metric is invariant under the simultaneous action of $\PSL(2,\CC)$ on $z_1$ and $z_2$. One can also describe $\calM$ as the quotient of $\PSL(2,\CC)$ by the stabilizer of the point $(0,\infty)$, that is, the complex subgroup generated by $\Lambda_0$.

We now construct the three embeddings together with some related structure. Each embedding is extended to a ``map of principal bundles'', which consists of compatible maps between their bases, total spaces, and structure groups. Let us begin with the map from the principal bundle $\tGG\to\HH^2$ to $\PSL(2,\CC)\to\calM$ and denote its constituent parts by $\zeta$, $J$, and $\omega_{\HH}$. The next equation includes the condition for compatibility between $\zeta$ and $J$ (expressed as a commutative diagram) as well as the definitions of these maps; $Z$ denotes the subgroup of $\tGG$ generated by $e^{2\pi\Lambda_0}$.
\begin{equation}
\begin{CD}
\tGG @>{J}>> \PSL(2,\CC)\\
@VVV @VVV\\
\HH^2 @>{\zeta}>> \calM
\end{CD}\qquad\quad
\begin{aligned}
J(g)&=g\quad (\text{reduced modulo $Z$}),\\[5pt]
\zeta(z)&=\bigl(z,\bar{z}^{\,-1}\bigr).
\end{aligned}
\end{equation}
The map $\omega_{\HH}$ (from the group of elements $h=e^{\theta\Lambda_0}$,\, $\theta\in\RR$ to such elements with complex $\theta$) should be a group homomorphism and satisfy the compatibility condition $J\circ\Ra(h)=\Ra(\omega_{\HH}(h))\circ J$, i.e.\ $J(gh^{-1})=J(g)\,\omega_{\HH}(h^{-1})$ for all $h$. Since $J$ is, essentially, trivial, such is $\omega_{\HH}$, namely, $\omega_{\HH}(h)=h$ (modulo $Z$). On the other hand, both principal bundle maps from $\tGG\to\widetilde{\AdS}_2$ to $\PSL(2,\CC)\to\calM$ involve this homomorphism of structure groups:
\begin{equation}
\omega_{\AdS}\bigl(e^{\theta\Lambda_2}\bigr)
=e^{i\theta\Lambda_0}\quad \text{for all }\, \theta\in\RR,\qquad
\text{i.e.}\quad
\omega_{\AdS}(h)=W^{-1}hW,
\end{equation}
where
\begin{equation}
W=e^{i(\pi/2)\Lambda_1}
=\frac{1}{\sqrt{2}}\begin{pmatrix} 1 & i \\ i & 1 \end{pmatrix},\qquad\quad
W(z)=\frac{z+i}{iz+1}.
\end{equation}
The other parts are defined below. Although $\tilde{\zeta}$ and $\mathring\zeta$ are not injective, they factor as the projection onto $\AdS_2$ followed by an embedding.
\begin{alignat}{2}
&\begin{CD}
\tGG @>{\tilde{J}}>> \PSL(2,\CC)\\
@VVV @VVV\\
\widetilde{\AdS}_2 @>{\tilde{\zeta}}>> \calM
\end{CD}\qquad\qquad&
&\begin{aligned}
\tilde{J}(g)&=gW,\\[5pt]
\tilde{\zeta}(\ph_1,\ph_2)&=\bigl(e^{i\ph_1},e^{i\ph_2}\bigr);
\end{aligned}
\displaybreak[0]\\[12pt]
&\begin{CD}
\tGG @>{\mathring{J}}>> \PSL(2,\CC)\\
@VVV @VVV\\
\widetilde{\AdS}_2 @>{\mathring{\zeta}}>> \calM
\end{CD}\qquad\qquad&
&\begin{aligned}
\mathring{J}(g)&=W^{-1}gW,\\[5pt]
\mathring{\zeta}(\ph_1,\ph_2)
&=\bigl(W^{-1}(e^{i\ph_1}),\,W^{-1}(e^{i\ph_2})\bigr)\\
&=\bigl(\tan(\pi/4-\ph_1/2),\,\tan(\pi/4-\ph_2/2)\bigr).
\end{aligned}
\end{alignat}
When the map $\zeta:\,\HH^2\to\calM$ is used together with $\tilde{\zeta}:\,\widetilde{\AdS}_2\to\calM$, the coordinates $z$, $\bar{z}$ on the hyperbolic plane correspond to the functions $z_1=e^{i\ph_1}$ and $z_2^{-1}=e^{-i\ph_2}$ on the anti-de Sitter space by the analytic continuation through $\calM$. If, on the other hand, $\widetilde{\AdS}_2$ is mapped to $\calM$ using $\mathring{\zeta}$, then the coordinate $\xi$ is consistent between $\HH^2$ and the Schwarzschild patch of $\widetilde{\AdS}_2$, and $\ph$ analytically continues to $it$.

\subsection{Definitions of spinors and two standard gauges}\label{sec_spinors}

Spinors on $\HH^2$, or any Riemannian surface, are associated with representations of the universal cover of $\SO(2)$, that is, the group $\tK$ generated by $\Lambda_0$. Let us consider the one-dimensional representation such that $\Lambda_0$ acts as the multiplication by $-i\nu$. Sections of the vector bundle associated with this representation and some principal $\tK$-bundle are called ``$\nu$-spinors''. More explicitly, a $\nu$-spinor is a function $\Psi$ from the total space of the principal bundle to the representation space (or simply the complex numbers) such that
\begin{equation}
(\Lambda_0^\ra-i\nu)\Psi=0.
\end{equation}
Here the superscript ``R'' refers to the action of the structure group on the total space. In the hyperbolic plane case, the principal $\tK$-bundle is given by the quotient map $\tGG\to\HH^2$, and the action in question is the R-action considered previously.

For calculational purposes, it is convenient to represent spinors by functions on the base space. This requires fixing a gauge, i.e.\ a cross section of the principal $\tK$-bundle. Let $s_{\HH}:\,\HH^2\to\tGG$ be such a cross section, and let $\psi(x)=\Psi(s_{\HH}(x))$ Any point of the fiber over $x\in\HH^2$ can be represented as $s_{\HH}(x)\,e^{-\theta\Lambda_0}$; hence,
\begin{equation}
\Psi\bigl(s_{\HH}(x)\,e^{-\theta\Lambda_0}\bigr)
=e^{-i\nu\theta}\psi(x).
\end{equation}
If $s_{\HH}(x)$ is replaced with $s_{\HH}(x)\,e^{-\tau(x)\Lambda_0}$, then $\psi(x)$ changes to $e^{-i\nu\tau(x)}\psi(x)$.

Spinors on $\widetilde{\AdS}_2$ are defined by the condition that $\Lambda_2$ acts as the multiplication by $\nu$. This definition is motivated by the relation between the structure group maps $\omega_{\AdS}$ and $\omega_{\HH}$; indeed, the generator $\Lambda_2$ in the anti-de Sitter case corresponds to $i\Lambda_0$ in the hyperbolic plane case. Thus,
\begin{equation}
\Psi\bigl(s_{\AdS}(x)\,e^{-\theta\Lambda_2}\bigr)
=e^{\nu\theta}\psi(x).
\end{equation}

Now, let us define cross sections $\tilde{s}_{\HH},\mathring{s}_{\HH}:\,\HH^2\to\tGG$ whose analytic continuations to $\calM$ are consistent with the respective embeddings of $\AdS_2$. We will give explicit formulas as well as some pictures. Cross sections of the principal bundle $\tGG\to\HH^2$ can be visualized as local frames on the unit disk. Indeed, consider the vector fields $\tilde{v}_0,\tilde{v}_1,\tilde{v}_2$ on $\tGG$ that correspond to the R-action of $\Lambda_0,\Lambda_1,\Lambda_2$. At each point $g=e^{\ph\Lambda_0}e^{\xi\Lambda_1}e^{-\vth\Lambda_0}$, they are given by the columns of the matrix $X^\ra(g)$, see Eq.~(\ref{XLR}). The first of them lies in the fiber and the other two are orthogonal to it. Projecting $\tilde{v}_1$ and $\tilde{v}_2$ on the base, we get these vectors $v_1$, $v_2$:
\begin{equation}
\begin{pmatrix}v_1^\xi\\[4pt] v_1^\ph\end{pmatrix}
=\begin{pmatrix}\cos\vth\\[4pt] -\dfrac{\sin\vth}{\sinh\xi}\end{pmatrix},
\qquad\quad
\begin{pmatrix}v_2^\xi\\[4pt] v_2^\ph\end{pmatrix}
=\begin{pmatrix}\sin\vth\\[4pt] \dfrac{\cos\vth}{\sinh\xi}\end{pmatrix}.
\end{equation}
Setting $g=s_{\HH}(x)$ gives an orthonormal frame at $x\in\HH^2$. The two particular cross sections are:
\begin{equation}
\tilde{s}_{\HH}(z)=e^{\ph\Lambda_0}e^{\xi\Lambda_1}\quad
\figbox{1.0}{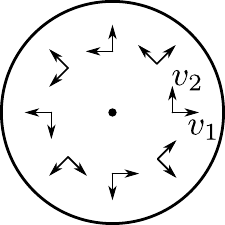}\qquad\qquad\quad
\mathring{s}_{\HH}(z)=e^{\ph\Lambda_0}e^{\xi\Lambda_1}e^{-\ph\Lambda_0}\quad
\figbox{1.0}{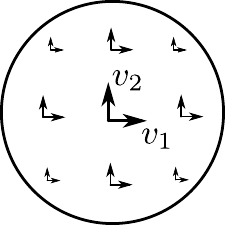}\qquad\qquad
\end{equation}
where $z=e^{i\ph}\tanh(\xi/2)$. The first one is multivalued because $(\xi,\ph)$ and $(\xi,\ph+2\pi)$ correspond to the same $z$ but $e^{\ph\Lambda_0}e^{\xi\Lambda_1}\not =e^{(\ph+2\pi)\Lambda_0}e^{\xi\Lambda_1}$.

As previously alluded to, $\tilde{s}_{\HH}$ is related to some cross section $\tilde{s}_{\AdS}$ of the principal bundle $\tGG\to\widetilde{\AdS}_2$. The correspondence between $\tilde{s}_{\HH}$, $\tilde{s}_{\AdS}$, and their common analytic continuation $\tilde{s}$ can be expressed by a commutative diagram and then translated to explicit equations:
\begin{equation}
\begin{CD}
\tGG @>{J}>> \PSL(2,\CC) @<{\tilde{J}}<< \tGG \\
@A{\tilde{s}_{\HH}}AA @A{\tilde{s}}AA @AA{\tilde{s}_{\AdS}}A \\
\HH^2 @>{\zeta}>> \calM @<{\tilde{\zeta}}<< \widetilde{\AdS}_2
\end{CD}\qquad\qquad
\begin{aligned}
\tilde{s}(z,\bar{z}^{\,-1})&=\tilde{s}_{\HH}(z),\\[5pt]
\tilde{s}(e^{i\ph_1},e^{i\ph_2})&=\tilde{s}_{\AdS}(\ph_1,\ph_2)\,W.
\end{aligned}
\end{equation}
(Although $\tilde{s}_{\HH}$ is multivalued and $\tilde{s}$ double-valued, $\tilde{s}_{\AdS}$ is well-defined.) The solution to the above equations is:
\begin{equation}
\tilde{s}_{\AdS}(\ph_1,\ph_2)=e^{\ph\Lambda_0}e^{\gamma\Lambda_1},\qquad
\text{where}\quad
\ph=\frac{\ph_1+\ph_2}{2},\quad\:
\gamma=-\ln\tan\frac{\ph_1-\ph_2}{4}.
\end{equation}
It is important that all the functions involved (namely, $\ph$ and $\gamma$) are real; that would not be the case if we began with $\mathring{s}_{\HH}$. The cross section $\mathring{s}_{\HH}$ is consistent with the other embedding of $\AdS_2$. Specifically,
\begin{equation}
\begin{CD}
\tGG @>{J}>> \PSL(2,\CC) @<{\mathring{J}}<< \tGG \\
@A{\mathring{s}_{\HH}}AA @A{\mathring{s}}AA @A{\mathring{s}_{\AdS}}AA \\
\HH^2 @>{\zeta}>> \calM @<{\mathring{\zeta}}<< \widetilde{\AdS}_2
\end{CD}\qquad\quad
\begin{aligned}
\mathring{s}(z,\bar{z}^{\,-1})&=\mathring{s}_{\HH}(z),\\[5pt]
\mathring{s}\bigl(W^{-1}(e^{i\ph_1}),\,W^{-1}(e^{i\ph_2})\bigr)
&=W^{-1}\mathring{s}_{\AdS}(\ph_1,\ph_2)\,W.
\end{aligned}
\end{equation}
Solving these equations requires slightly more work. The result is this:
\begin{equation}
\mathring{s}_{\AdS}(\ph_1,\ph_2)
=\tilde{s}_{\AdS}(\ph_1,\ph_2)\,e^{-\tau\Lambda_2},\qquad
\text{where}\quad
\tau=\ln\frac{\sin(\pi/4+\ph_1/2)}{\sin(\pi/4-\ph_2/2)}.
\end{equation}

Spinors written in the ``tilde gauge'' (i.e.\ using $\tilde{s}_{\HH}$ or $\tilde{s}_{\AdS}$) and in the ``disk gauge'' (using $\mathring{s}_{\HH}$ or $\mathring{s}_{\AdS}$) are related as follows:
\begin{equation}
\wideboxed{
\mathring{\psi}^{\HH}(z)
=\bigl(z/\bar{z}\bigr)^{-\nu/2}\,\tilde{\psi}^{\HH}(z),\qquad
\mathring{\psi}^{\AdS}(\ph_1,\ph_2)
=\biggl(\frac{\sin(\pi/4+\ph_1/2)}{\sin(\pi/4-\ph_2/2)}\biggr)^{\nu}
\tilde{\psi}^{\AdS}(\ph_1,\ph_2)
}
\end{equation}
In either gauge, the analytic continuation of spinors is similar to that of ordinary functions. It still depends on the embedding of $\AdS_2$ in $\calM$, therefore the explicit expressions are different:
\begin{align}
\tilde{\psi}^{\HH}(z)
&=\tilde{\psi}\bigl(z,\bar{z}^{\,-1}\bigr), &
\tilde{\psi}^{\AdS}(\ph_1,\ph_2)
&=\tilde{\psi}\bigl(e^{i\ph_1},e^{i\ph_2}\bigr),\\[3pt]
\mathring{\psi}^{\HH}(z)
&=\mathring{\psi}\bigl(z,\bar{z}^{\,-1}\bigr), &
\mathring{\psi}^{\AdS}(\ph_1,\ph_2)
&=\mathring{\psi}\bigl(\tan(\pi/4-\ph_1/2),\,\tan(\pi/4-\ph_2/2)\bigr).
\end{align}

\subsection{Spinors on $\HH^2$}

\paragraph{Analytic spinors:} Many applications involve $\nu$-spinors that can be expressed by a convergent Taylor series in $z_1=z$ and $z_2^{-1}=\bar{z}$ for $|z|,|\bar{z}|<1$. Such spinors transform under maps $V\in\tGG$ as follows:
\begin{equation}
(V\mathring{\psi})(z_1,z_2)=\biggl(\frac{dz_1}{dw_1}\biggr)^{-\nu/2}
\biggl(\frac{dz_2^{-1}}{dw_2^{-1}}\biggr)^{\nu/2}\mathring{\psi}(w_1,w_2)
\qquad\text{for}\quad z_1=V(w_1),\quad z_2=V(w_2).
\end{equation}
This is a special case of the transformation rule for holomorphic $(\lambda_1,\lambda_2)$-forms, i.e.\ functions of $z_1$, $z_2$ that transform as elements of $\calF_{\lambda_1}^{+}$ in the first variable and $\calF_{\lambda_2}^{-}$ in the second variable, cf.\ Eq.~(\ref{aftrans}). Such forms are written symbolically as
\begin{equation}
f=\of(z_1,z_2)\,(-i\,dz_1)^{\lambda_1}(i\,dz_2^{-1})^{\lambda_2}.
\end{equation}
For example, $(1-z_1/z_2)^{-2\lambda}(dz_1)^\lambda(dz_2^{-1})^{\lambda}$ is a $(\lambda,\lambda)$-form that is invariant under $\tGG$ (and more general linear fractional maps in a suitable domain). Thus, $\nu$-spinors are $\bigl(\frac{\nu}{2},-\frac{\nu}{2}\bigr)$-forms with respect to maps $V\in\tGG$. Conversely, any $(\lambda_1,\lambda_2)$-form $f$ can be written as
\begin{equation}
\of(z_1,z_2)=\bigl(1-z_1/z_2\bigr)^{-(\lambda_1+\lambda_2)}\,
\mathring{\psi}(z_1,z_2),
\end{equation}
where $\psi$ is a $(\lambda_1-\lambda_2)$-spinor. 

\begin{figure}
\centering\(\displaystyle
\begin{array}{c@{\qquad}c@{\qquad}c}
\figbox{1.0}{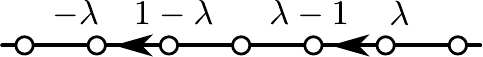} & \figbox{1.0}{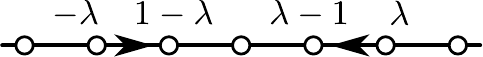} & \figbox{1.0}{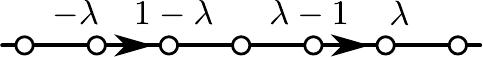}
\vspace{8pt}\\
\nu\le-\lambda & -\lambda<\nu<\lambda & \nu\ge\lambda
\end{array}
\)
\caption{The action of $L_{-1}$, $L_{1}$ on Casimir eigenfunctions for $\lambda=1,\frac{3}{2},2,\ldots$ and $\nu\in\lambda+\ZZ$. A circle with label $m\in\nu+\ZZ$ represents the basis function $\psie{\lambda}{\nu,+}{m}$ if $m\ge\nu$ and $\psie{\lambda}{\nu,-}{m}$ if $m\le\nu$.}
\label{fig_casef}
\end{figure}

\paragraph{Casimir eigenfunctions:} Among $\nu$-spinors, let us consider the common eigenfunctions of the Casimir operator with the eigenvalue $\lambda(1-\lambda)$ and the operator $L_0$ with the eigenvalue $-m$. Since spinors on $\HH^2$ are a certain type of functions on $\tGG$ on which the group operates by the L-action, we can simply use the equations from Section~\ref{sec_Casimir_ef} with $l=m$,\,\, $r=-\nu$,\,\, $u=z\bar{z}=z_1/z_2$, and $e^{i\ph}=\sqrt{z/\bar{z}}=\sqrt{z_1z_2}$. In particular, the eigenfunctions $e^{i(l\ph+r\vth)}A_{\lambda,l,r}(u)$ and $e^{i(l\ph+r\vth)}A_{\lambda,-l,-r}(u)$ in the disk gauge (i.e.\ with $\vth=\ph$) become
\begin{equation}\label{psi_lnm}
\wideboxed{
\begin{aligned}
\matelem{\mathring{\psi}}{\lambda}{\nu,+}{m}(z_1,z_2)
&=z_1^{m-\nu}(1-z_1/z_2)^{\lambda}\,
\hgfs\bigl(\lambda+m,\,\lambda-\nu,\,1+m-\nu;\,z_1/z_2\bigr)
\\[5pt]
\matelem{\mathring{\psi}}{\lambda}{\nu,-}{m}(z_1,z_2)
&=z_2^{m-\nu}(1-z_1/z_2)^{\lambda}\,
\hgfs\bigl(\lambda-m,\,\lambda+\nu,\,1-m+\nu;\,z_1/z_2\bigr)
\end{aligned}
}
\end{equation}
These functions are nonsingular if and only if $m\in\nu+\ZZ$, in which case they are linearly dependent. Choosing one of them that is nonzero for each given $m$, e.g.\ $\psie{\lambda}{\nu,+}{m}$ for $m\ge\nu$ and $\psie{\lambda}{\nu,-}{m}$ for $m\le\nu$, we obtain a basis of some representation of $\tGG$. The group action on the basis functions is characterized by the equations below and illustrated by Figure~\ref{fig_casef}.
\begin{equation}
\begin{alignedat}{2}
L_{-1}\psie{\lambda}{\nu,+}{m}
&=-\psie{\lambda}{\nu,+}{m-1}, \qquad&
L_{1} \psie{\lambda}{\nu,+}{m}
&=-(m+\lambda)(m+1-\lambda)\,\psie{\lambda}{\nu,+}{m+1},
\\[3pt]
L_{-1}\psie{\lambda}{\nu,-}{m}
&=(m-\lambda)(m-1+\lambda)\,\psie{\lambda}{\nu,-}{m-1}, \qquad&
L_{1} \psie{\lambda}{\nu,-}{m}
&=\psie{\lambda}{\nu,-}{m+1}.
\end{alignedat}
\end{equation}

\paragraph{Intertwiner from the space $\calF^{\nu}_{1-\lambda}$ to $\nu$-spinors on the hyperbolic plane:}
A nontrivial intertwiner $E^{\nu}_{\lambda}$ of this type exists and is unique up to an overall factor.\footnote{In quantum holography, such an intertwiner is interpreted as a bulk-boundary propagator, where $\lambda$ is the scaling dimension of the field from the boundary point of view.} Its action on the basis vectors $f_{1-\lambda,m}\in\calF^{\nu}_{1-\lambda}$ is given by the equation
\begin{equation}
E^{\nu}_{\lambda}f_{1-\lambda,m}
=\frac{\Gamma(\lambda+m)}{\Gamma(\lambda+\nu)}\,\psie{\lambda}{\nu,+}{m}
=\frac{\Gamma(\lambda-m)}{\Gamma(\lambda-\nu)}\,\psie{\lambda}{\nu,-}{m}
\qquad\text{for }\, m\in\nu+\ZZ.
\end{equation}
If we regard the $\nu$-spinor $E^{\nu}_{\lambda}f_{1-\lambda,m}$ as a function of $g\in\tGG$, then
\begin{equation}
\bigl(E^{\nu}_{\lambda}f_{1-\lambda,m}\bigr)(g)
=\Bigl(f_{\lambda,-\nu},\,F^{\nu}_{1-\lambda}(g^{-1})f_{1-\lambda,m}\Bigr).
\end{equation}
where $F^{\nu}_{1-\lambda}(g^{-1})$ is the action of the group element $g^{-1}$ in the representation space $\calF^{\nu}_{1-\lambda}$ and the big parentheses denote the integral of the product of two functions with $\frac{d\ph}{2\pi}$. While this integral can be calculated directly, we note that the right-hand side of the above equation is a non-unitary version of the matrix element functions considered in Section~\ref{sec_Fourier1}. Such functions are transformed as the basis vectors $f_{1-\lambda,m}\in\calF^{\nu}_{1-\lambda}$, which is exactly the intertwiner property.

Let us also write the $\nu$-spinor $E^{\nu}_{\lambda}f$ for an arbitrary $f\in\calF^{\nu}_{1-\lambda}$ in the disk gauge:
\begin{equation}
\wideboxed{
\bigl(E^\nu_{\lambda}f\bigr)^{\!\hbox{\tiny$\circ$}}(z)
=\int_{0}^{2\pi}
\frac{(1-z\bar{z})^{\lambda}}
{(1-\bar{z}e^{i\ph})^{\lambda-\nu}\,(1-ze^{-i\ph})^{\lambda+\nu}}\:
e^{-i\nu\ph}\tilde{f}(\ph)\,\frac{d\ph}{2\pi}
}
\end{equation}
This equation is proved by expanding the integrand in $z$ and $\bar{z}$. Once again, there is an independent argument showing that the integral operator on the right-hand side defines an intertwiner. Indeed, its kernel function corresponds to the $\tGG$-invariant form
\begin{equation}
\frac{(1-z_1/z_2)^{\lambda}}
{(1-e^{i\ph}/z_2)^{\lambda-\nu}\,(1-z_1/e^{i\ph})^{\lambda+\nu}}\:
(-i\,dz_1)^{\nu/2}(i\,dz_2^{-1})^{-\nu/2}(d\ph)^{\lambda},
\qquad\quad
z_1=z,\quad z_2=\bar{z}^{\,-1}.
\end{equation}

\begin{figure}
\centering\(\figbox{1.0}{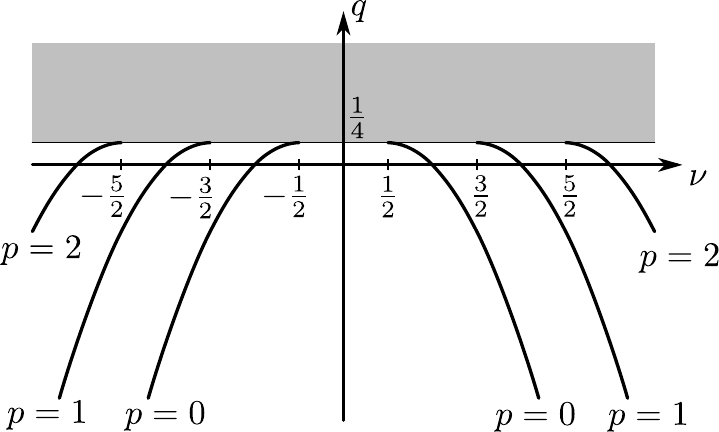}\hspace{3cm}\figbox{1.0}{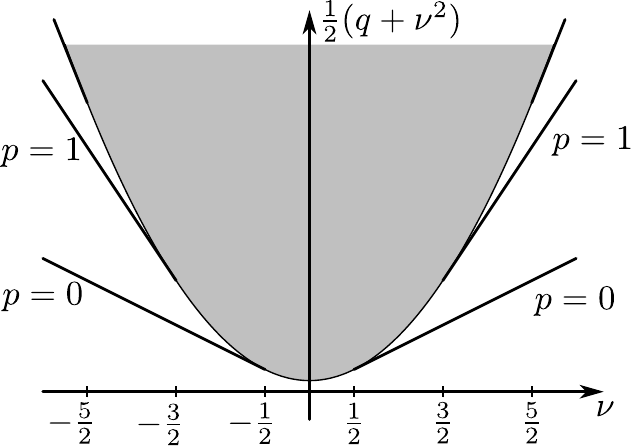}\)
\caption{The spectra of the operators $Q$ and $\frac{1}{2}(Q+\nu^2)$ as functions of the spin value $\nu$.}
\label{fig_spectrum}
\end{figure}

\paragraph{Square-integrable spinors:}
A basis in the Hilbert space $\calH_{\HH}^{\nu}$ of square-integrable $\nu$-spinors consists of the matrix element functions $\Um{\lambda}{\nu}{m}(g)$, see equation~(\ref{U_lnm}). They can also be written in terms of the variables $(z_1,z_2)$ using the new notation $\psie{\lambda}{\nu,\pm}{m}$:
\begin{equation}\label{psi_symm}
\psie{\lambda}{\nu}{m}=
\sqrt{\frac{\Gamma(\lambda+m)\,\Gamma(1-\lambda+m)}
{\Gamma(\lambda+\nu)\,\Gamma(1-\lambda+\nu)}}\, \psie{\lambda}{\nu,+}{m}
=(-1)^{m-\nu} \sqrt{\frac{\Gamma(\lambda-m)\,\Gamma(1-\lambda-m)}
{\Gamma(\lambda-\nu)\,\Gamma(1-\lambda-\nu)}}\, \psie{\lambda}{\nu,-}{m}.
\end{equation}
These functions transform as the basis vectors $\ket{m}$ of the principal series representation $\calC^{\nu}_{1/4+s^2}$ with $\lambda=\frac{1}{2}+is$ and $m\in\nu+\ZZ$, or the discrete series representation $\calD^{+}_{\lambda}$ with $\lambda>\frac{1}{2}$ and $\nu,m\in\{\lambda,\lambda+1,\ldots\}$, or the representation $\calD^{-}_{\lambda}$ with $\lambda>\frac{1}{2}$ and $\nu,m\in\{-\lambda,-\lambda-1,\ldots\}$. Thus, $\calH_{\HH}^{\nu}$ splits into the indicated irreps, each of which enters with multiplicity $1$. Symbolically,
\begin{equation}\label{spec_H2}
\calH_{\HH}^\nu\cong\int_{0}^{\infty}\!\calC^{\nu}_{1/4+s^2}\,ds
\oplus{}\bigoplus_{\lambda}\calD_{\lambda}^{\sgn\nu},\qquad
\text{where}\quad
\lambda=|\nu|-p>\frac{1}{2},\quad p\in\{0,1,2,\ldots\}.
\end{equation}
The spectrum of the Casimir operator $Q$, i.e.\ the set of numbers $q=\frac{1}{4}+s^2$ and $q=\lambda(1-\lambda)$ in the above equation, is plotted in Figure~\ref{fig_spectrum}.

As an aside, the qualitative form of the spectrum has an interesting physical interpretation. An operator closely related to $Q$, namely, $-\frac{1}{2}\nabla^2=\frac{1}{2}(Q+\nu^2)$ describes a quantum particle with spin $\nu$ on the hyperbolic plane. It may also be viewed as the Hamiltonian of a nonrelativistic spinless particle with unit mass and electric charge in a magnetic field of strength $\nu$. In the flat geometry, such a Hamiltonian has a purely discrete spectrum, the Landau levels. This is because all classical trajectories are closed. However, the trajectory of a charged particle on the hyperbolic plane can be either closed or open, depending on the ratio between the velocity and magnetic field:
\begin{equation}
\left|\frac{v}{\nu}\right|<1:\quad \figbox{1.0}{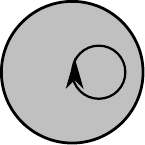}\qquad\qquad\quad
\left|\frac{v}{\nu}\right|>1:\quad \figbox{1.0}{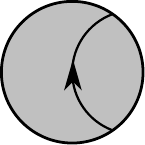}
\end{equation}
Therefore, for velocities $v>\nu$, i.e.\ energies greater than $\nu^2/2$, the spectrum becomes continuous. Of course, this argument is very rough and does not give the exact spectrum.

Finally, let us consider the decomposition of the unit operator of the Hilbert space $\calH_{\HH}^{\nu}$ into the projectors onto its irreducible components:
\begin{empheq}[box=\widebox]{gather}
\label{decomp_id}
\mathbf{1}=(2\pi)^{-1}\biggl(
\int_{0}^{\infty} ds\, 
\frac{s\,\sinh(2\pi s)}{\cosh(2\pi s)+\cos(2\pi\nu)}\, \Pi^{\nu}_{1/2+is}
+\sum_{\substack{\lambda=|\nu|-p>1/2\\ p=0,1,2,\ldots}}
\bigl(\lambda-\tfrac{1}{2}\bigr)\, \Pi^{\nu}_{\lambda}
\biggr)\\[5pt]
\label{Pi_ln_def}
\Pi^{\nu}_{\lambda}
=\sum_{m}\ket{\psie{\lambda}{\nu}{m}} \bra{\psie{\lambda}{\nu}{m}},\qquad
m\in\nu+\ZZ\quad (\text{restricted for discrete series})
\end{empheq}
Here we have used the Plancherel measure~(\ref{Plancherel}) together with the constraint $\nu-\mu\in\ZZ$. This is a slightly more detailed explanation. The basis functions $\ket{\psie{\lambda}{\nu}{m}}$ are, essentially, the same as $\ket{\Um{\lambda}{\nu}{m}}$, which are related to the Plancherel measure. The constraints on $\lambda$, $\nu$, $m$ in the above equations follow from those in Eqs.~(\ref{ortho_discr}), (\ref{ortho_cont}). However, in the first case the inner product is defined as an integral over $\HH^2$, and in the second over $\tGG$. Therefore,
\begin{equation}
\bbraket{\Um{\lambda}{\nu}{m}}{\Um{\lambda'}{\nu'}{m'}}
=2\pi\,\delta(\nu-\nu')\,
\bbraket{\psie{\lambda}{\nu}{m}}{\psie{\lambda'}{\nu}{m'}}.
\end{equation}
The relation between the decomposition measures is the inverse one, that is, equation~(\ref{decomp_id}) uses the Plancherel measure multiplied by $2\pi\sum_{n\in\ZZ}\delta(\nu-\mu-n)$.

The projectors $\Pi^{\nu}_{\lambda}$ are integral operators with the kernel functions that depend on $z,w\in\HH^2$ and the integration measure $4(1-w\bar{w})^{-2}\,dw\,d\bar{w}$. In terms of the variables $z_1=z$,\, $z_2=\bar{z}^{\,-1}$,\, $w_1=w$,\, $w_2=\bar{w}^{-1}$, the kernel function is
\begin{equation}
\mathring{\Pi}^{\nu}_{\lambda}(z_1,z_2;w_1,w_2)
=\sum_{m} (-1)^{m-\nu}\, \matelem{\mathring{\psi}}{\lambda}{\nu}{m}(z_1,z_2)\,
\matelem{\mathring{\psi}}{\lambda}{-\nu}{-m}(w_1,w_2),
\end{equation}
If $(w_1,w_2)=(0,\infty)$, then only the $m=\nu$ term in the sum is nonzero:
\begin{equation}
\mathring{\Pi}^{\nu}_{\lambda}(z_1,z_2;0,\infty)
=\matelem{\mathring{\psi}}{\lambda}{\nu}{\nu}(z_1,z_2)
=(1-z_1/z_2)^{\lambda}\,
\hgfs\bigl(\lambda+\nu,\,\lambda-\nu,\,1;\,z_1/z_2\bigr).
\end{equation}
The general case is reduced to this one using a symmetry argument. Indeed, the kernel function defines a form of degree $\bigl(\frac{\nu}{2},-\frac{\nu}{2}\bigr)$ with respect to $(z_1,z_2)$ and $\bigl(-\frac{\nu}{2},\frac{\nu}{2}\bigr)$ with respect to $(w_1,w_2)$ when both pairs of variables are close to $(0,\infty)$. This form is invariant under linear fractional maps because the projector commutes with the $\sL_2$ action. The map
\begin{equation}
V:\,z\mapsto\frac{z-w_1}{1-z/w_2}
\end{equation}
sends $w_1$ to $0$ and $w_2$ to $\infty$. Therefore, $\mathring{\Pi}^{\nu}_{\lambda}(z_1,z_2;w_1,w_2)$ can be written as an arbitrary invariant form of the same type multiplied by some ordinary function of $V(z_1)$ and $V(z_2)$. More concretely,
\begin{equation}
\mathring{\Pi}^{\nu}_{\lambda}(z_1,z_2;w_1,w_2)
=(1-z_1/w_2)^{-\nu}(1-w_1/z_2)^{\nu}\,
\mathring{\Pi}^{\nu}_{\lambda}\bigl(V(z_1),V(z_2);\,0,\infty\bigr).
\end{equation}
Thus,
\begin{empheq}[box=\widebox]{gather}
\label{spinor_proj}
\mathring{\Pi}^{\nu}_{\lambda}(z_1,z_2;w_1,w_2)
=\biggl(\frac{1-w_1/z_2}{1-z_1/w_2}\biggr)^{\nu} (1-u)^{\lambda}\,
\hgfs\bigl(\lambda+\nu,\,\lambda-\nu,\,1;\,u\bigr),\qquad
\\[5pt]
\label{u_in_proj}
\text{where}\qquad u=\frac{(z_1-w_1)(z_2-w_2)}{(z_1-w_2)(z_2-w_1)}
\end{empheq}

\section{Tensor products of unitary irreps}

A standard problem in representation theory is to decompose the product of two unitary irreps into irreps with multiplicities. For the group $\SL(2,\RR)$, this task was accomplished by Repka~\cite{Repka76,Repka78}. We will not attempt to give a complete solution for $\tGG$. Rather, we will sketch a general recipe and work out the discrete series cases, $\calD^{\pm}_{\lambda_1}\otimes\calD^{\pm}_{\lambda_2}$, which are relatively simple and relevant to the SYK model.

The problem of splitting the representation $\calU_{\lambda_1}^{\mu_1}\otimes\calU_{\lambda_2}^{\mu_2}$ is equivalent to finding all intertwiners
\begin{equation}
\Upsilon:\, \calU_{\lambda}^{\mu}\to
\calU_{\lambda_1}^{\mu_1}\otimes\calU_{\lambda_2}^{\mu_2},\qquad\quad
\mu=\mu_1+\mu_2
\end{equation}
and selecting those that are normalizable or $\delta$-normalizable with respect to $\lambda$. The matrix elements $\bra{m_1,m_2}\Upsilon\ket{m}$ will be called ``Clebsch-Gordan coefficients''. For now, let us not worry about normalizability and discuss a related task: find all $\tGG$-invariant forms
\begin{equation}\label{generating_form}
\tilde{Y}\bigl(e^{i\ph_1},e^{i\ph_2},e^{i\ph_3}\bigr)\,
(d\ph_1)^{\lambda_1}(d\ph_2)^{\lambda_2}(d\ph_3)^{\lambda_3},
\end{equation}
where, $e^{i\ph_1}$, $e^{i\ph_2}$, $e^{i\ph_3}$ represent three points on the unit circle. Choosing some numbers $\mu_1$, $\mu_2$, $\mu_3$ with zero sum and using the twisted periodicity conditions, we can extend $\tilde{Y}$ from the fundamental domain $2\pi+\ph_3>\ph_1,\ph_2>\ph_3$ to a function of real variables $\ph_1$, $\ph_2$, $\ph_3$. It is understood as a generalized function and may be defined by the Fourier expansion
\begin{equation}\label{genfun_Fourier}
\tilde{Y}(z_1,z_2,z_3)
=\sum_{\substack{m_1\in\mu_1+\ZZ,\,\,m_2\in\mu_2+\ZZ\\ m_1+m_2+m_3=0}}
C_{m_1,m_2,m_3}\,z_1^{m_1}z_2^{m_2}z_3^{m_3}
\end{equation}
with the coefficients $C_{m_1,m_2,m_3}$ growing at most polynomially. To express the Clebsch-Gordan coefficients, let $\lambda_3=\lambda$,\, $\mu_3=-\mu$, and let us consider $\tilde{Y}$ as the integral kernel of the intertwiner
\begin{equation}
Y:\,\calF_{1-\lambda}^{\mu}\to
\calF_{\lambda_1}^{\mu_1}\otimes\calF_{\lambda_2}^{\mu_2},\qquad\quad
Y=\bigl(\Xi_{\lambda_1\spinup}^{\mu_1\pm}
\otimes\Xi_{\lambda_2\spinup}^{\mu_2\pm}\bigr)\,
\Upsilon\,\Xi_{\lambda\spindown}^{\mu\pm}.
\end{equation}
Here $\Xi_{\lambda\spindown}^{\mu\pm}$, $\Xi_{\lambda\spinup}^{\mu\pm}$ are defined by equation~(\ref{Xiupdown}) and each of the three signs is individually chosen. (This choice only matters when the corresponding irrep belongs to a discrete series.) Thus,
\begin{equation}\label{genfun_coeff}
C_{m_1,m_2,-m}=
c^{\pm}_{\lambda_1,m_1}c^{\pm}_{\lambda_2,m_2}c^{\pm}_{\lambda,m}\,
\bra{m_1,m_2}\Upsilon\ket{m}.
\end{equation}

The invariance of the generating function $\tilde{Y}$ under $L_{1}$ and $L_{-1}$ is expressed as linear relations between $C_{m_1-1,m_2,m_3}$, $C_{m_1,m_2-1,m_3}$, $C_{m_1,m_2,m_3-1}$ and between $C_{m_1+1,m_2,m_3}$, $C_{m_1,m_2+1,m_3}$, $C_{m_1,m_2,m_3+1}$. In general, this system of equations has multiple linearly independent solutions. But we are considering only those values of $\lambda_j$, $m_j$ that correspond to unitary irreps. With this restriction, the linear relations in the allowed region of $(m_1,m_2,m_3)$ are nondegenerate and can be turned into recurrences, which are solved beginning with just two Fourier coefficients. Thus, the solution space is at most two-dimensional. Its general form is easy to guess:
\begin{equation}\label{genfun1}
\tilde{Y}\bigl(e^{i\ph_1},e^{i\ph_2},e^{i\ph_3}\bigr)
=a\,|\ph_{12}|^{-\lambda_1-\lambda_2+\lambda_3}
|\ph_{13}|^{-\lambda_1+\lambda_2-\lambda_3}
|\ph_{23}|^{\lambda_1-\lambda_2-\lambda_3},\qquad
\ph_{jk}=2\sin\frac{\ph_j-\ph_k}{2},
\end{equation}
where $a$ takes on two different values depending on the cyclic order of $\ph_1$, $\ph_2$, $\ph_3$. However, this expression might require regularization when two or three points coincide. The problem arises if any of the singularities is non-integrable, that is, if one of the exponents $-\lambda_1-\lambda_2+\lambda_3$,\,\, $-\lambda_1+\lambda_2-\lambda_3$,\,\, $\lambda_1-\lambda_2-\lambda_3$ has real part less than or equal to $-1$ or if the real part of the sum of all three exponents is less than or equal to $-2$. This can only happen if a discrete series representation is involved. But if, say, $\calU_{\lambda_1}^{\mu_1}=\calD_{\lambda_1}^{+}$, then the generating form~(\ref{generating_form}) is holomorphic in $z_1=e^{i\ph_1}$ for $|z_1|<1$. In this case, the regularization is achieved by analytic continuation. Since both cyclic orders are just limiting cases of $z_1$ being inside the circle, the intertwiner space is one-dimensional.

\subsection{$\calD^{+}_{\lambda_1}\otimes\calD^{+}_{\lambda_2}$}

By analogy with $\SU(2)$ representations, it is clear that
\begin{equation}
\calD^{+}_{\lambda_1}\otimes\calD^{+}_{\lambda_2}
\cong\bigoplus_{n=0}^{\infty}\calD^{+}_{\lambda_1+\lambda_2+n}.
\end{equation}
For each given $n$, there is a unique (up to an overall factor) intertwiner
\begin{equation}
\Upsilon^{++}_{\lambda_1,\lambda_2;\lambda}:\,
\calD^{+}_{\lambda}\to\calD^{+}_{\lambda_1}\otimes\calD^{+}_{\lambda_2},\qquad
\lambda=\lambda_1+\lambda_2+n.
\end{equation}
Its generating function~(\ref{genfun_Fourier}) with all three signs in~(\ref{genfun_coeff}) set to ``$+$'' is
\begin{equation}
\tilde{Y}^{++\,+}_{\lambda_1,\lambda_2,\lambda}(z_1,z_2,w)
=z_1^{\lambda_1}z_2^{\lambda_2}w^{-\lambda}\,
(z_2-z_1)^{n}(1-z_1/w)^{-2\lambda_1-n}(1-z_2/w)^{-2\lambda_2-n}
\end{equation}

The rest of this subsection is concerned with the decomposition of identity for $\calD^{+}_{\lambda_1}\otimes\calD^{+}_{\lambda_2}$. First, we define partial generating functions as the Taylor coefficients with respect to $w^{-1}$, excluding the $c^{+}_{\lambda,m}$ factor:
\begin{equation}
\tilde{Y}^{++}_{\lambda_1,\lambda_2;\lambda,m}(z_1,z_2)
=\sum_{m_1,m_2} c^{+}_{\lambda_1,m_1}c^{+}_{\lambda_2,m_1}\,
\bbra{m_1,m_2}\Upsilon^{++}_{\lambda_1,\lambda_2;\lambda}\bket{m}\,
z_1^{m_1}z_2^{m_2}.
\end{equation}
Recall that $c^{+}_{\alpha,\alpha+k}=\sqrt{\Gamma(2\alpha+k)/k!}$. A straightforward calculation (where we use the Pochhammer symbol, $(\alpha)_r=\alpha\cdots(\alpha+r-1)$) shows that
\begin{equation}
\label{partgf1}
\begin{aligned}
\tilde{Y}^{++}_{\lambda_1,\lambda_2;\,\lambda,\,\lambda+k}(z_1,z_2)
&=\sqrt{\frac{k!}{\Gamma(2\lambda+k)}}\,
z_1^{\lambda_1}z_2^{\lambda_2}(z_2-z_1)^{n}\,
\sum_{k_1+k_2=k}^{k}\frac{(2\lambda_1+n)_{k_1}}{k_1!}\,
\frac{(2\lambda_2+n)_{k_2}}{k_2!}\,z_1^{k_1}z_2^{k_2}\\[3pt]
&=\sqrt{\frac{\Gamma(2\lambda+k)}{k!}}\,
z_1^{\lambda_1}z_2^{\lambda_2+n+k}\,
(1-z_1/z_2)^n\,\hgfs\bigl(-k,\,2\lambda_1+n,\,2\lambda;\,1-z_1/z_2\bigr).
\end{aligned}
\end{equation}
For example, $\tilde{Y}^{++}_{\lambda_1,\lambda_2;\lambda,\lambda}(z_1,z_2) =\Gamma(2\lambda)^{-1/2}z_1^{\lambda_1}z_2^{\lambda_2}(z_2-z_1)^{n}$. Summing up the squares of the Clebsch-Gordan coefficients in this special case, we find the norm of the intertwiner:
\begin{equation}
(\Upsilon^{++}_{\lambda_1,\lambda_2;\lambda})^{\dag}\,
\Upsilon^{++}_{\lambda_1,\lambda_2;\lambda}
=\frac{n!}{(2\lambda-1)\,\Gamma(2\lambda_1+2\lambda_2+n-1)\,
\Gamma(2\lambda_1+n)\,\Gamma(2\lambda_2+n)}\,
\mathbf{1}_{\lambda}^{+}.
\end{equation}
Thus,
\begin{empheq}[box=\widebox]{gather}
\mathbf{1}_{\lambda_1}^{+}\otimes\mathbf{1}_{\lambda_2}^{+}
=\!\sum_{\substack{\lambda=\lambda_1+\lambda_2+n\\[1pt] n=0,1,2,\ldots}}\!\!\!
\frac{(2\lambda-1)\,\Gamma(\lambda-1+\lambda_1+\lambda_2)\,
\Gamma(\lambda+\lambda_1-\lambda_2)\,\Gamma(\lambda-\lambda_1+\lambda_2)}{n!}\,
\Pi^{++}_{\lambda_1,\lambda_2;\,\lambda}
\\[3pt]
\Pi^{++}_{\lambda_1,\lambda_2;\lambda}
=\Upsilon^{++}_{\lambda_1,\lambda_2;\lambda}
(\Upsilon^{++}_{\lambda_1,\lambda_2;\lambda})^{\dag}
\end{empheq}
The generating function for the (unnormalized) projector $\Pi^{++}_{\lambda_1,\lambda_2;\lambda}$,
\begin{equation}
\tilde{\Pi}^{++}_{\lambda_1,\lambda_2;\lambda}(z_1,z_2;w_1,w_2)
=\sum_{m}\tilde{Y}^{++}_{\lambda_1,\lambda_2;\lambda,m}(z_1,z_2)\,
\tilde{Y}^{++}_{\lambda_1,\lambda_2;\lambda,m}(w_1^{-1},w_2^{-1}),
\end{equation}
is calculated by analogy with spinors. We first assume that $z_2\to 0$ and $w_1\to\infty$, so that only the $(k_1,k_2)=(k,0)$ terms in the expression for  $\tilde{Y}^{++}_{\lambda_1,\lambda_2;\lambda,\lambda+k}(z_1,z_2)$ and the $(k_1,k_2)=(0,k)$ terms in $\tilde{Y}^{++}_{\lambda_1,\lambda_2;\lambda,\lambda+k}(w_1^{-1},w_2^{-1})$ are present (see Eq.~(\ref{partgf1})). The general case is reduced to this one using symmetry. The result is as follows, where $u$ is defined by equation~(\ref{u_in_proj}):
\begin{empheq}[box=\widebox]{gather}
\tilde{\Pi}^{++}_{\lambda_1,\lambda_2;\lambda}(z_1,z_2;w_1,w_2)
=\frac{(z_1/w_1)^{\lambda_1}(z_2/w_2)^{\lambda_2}}
{(1-z_1/w_1)^{2\lambda_1}(1-z_2/w_2)^{2\lambda_2}}\,
(-\chi)^n\, \hgfs\bigl(2\lambda_1+n,\,2\lambda_2+n,\,2\lambda;\,\chi\bigr)
\\[6pt]
\lambda=\lambda_1+\lambda_2+n,\qquad\quad
\chi=1-u^{-1}=\frac{(z_1-z_2)(w_1-w_2)}{(z_1-w_1)(z_2-w_2)}
\end{empheq}

\subsection{$\calD^{+}_{\lambda_1}\otimes\calD^{-}_{\lambda_2}$}

The space $\calD^{+}_{\lambda_1}\otimes\calD^{-}_{\lambda_2}$ maps onto $\calF^{+}_{\lambda_1}\otimes\calF^{-}_{\lambda_2}$. The latter consists of holomorphic $\nu$-spinors with $\nu=\lambda_1-\lambda_2$. However, the norm on the original space differs from the spinor norm defined by the integral over $\HH^2$. One could use some functional analysis to characterize the relation between the Hilbert spaces $\calD^{+}_{\lambda_1}\otimes\calD^{-}_{\lambda_2}$ and $\calH^\nu_{\HH}$, cf.\ Proposition~7.2 in~\cite{Repka78}. We instead directly construct the irreducible decomposition
\begin{equation}\label{Dpm_space_decomp}
\calD^{+}_{\lambda_1}\otimes\calD^{-}_{\lambda_2}
\cong\int_{0}^{\infty}\!\calC^{\nu}_{1/4+s^2}\,ds
\oplus{}\bigoplus_{\substack{\lambda=|\nu|-p>1/2\\ p=0,1,2,\ldots}}\!\!
\calD_{\lambda}^{\sgn\nu}
\oplus\Bigl(
\calC^{\nu}_{\lambda(1-\lambda)}\,
\text{ for }\lambda=\lambda_1+\lambda_2<\tfrac{1}{2}\Bigr)
\end{equation}
by analogy with the derivation of Eq.~(\ref{spec_H2}). In doing so, we reuse the Casimir eigenfunctions as partial generating functions, but calculate their norms using the inner product on $\calD^{+}_{\lambda_1}\otimes\calD^{-}_{\lambda_2}$. The key observation is that the asymptotics of the spinors and the Clebsch-Gordan coefficients are closely related, and the corresponding integral and sum converge for the same values of $\lambda$ (with one exception that results in the extra term in Eq.~(\ref{Dpm_space_decomp})).

According to the general scheme, we consider an intertwiner
\begin{equation}
\Upsilon^{+-}_{\lambda_1,\lambda_2;\lambda}:\,
\calU^{\nu}_{\lambda}\to\calD^{+}_{\lambda_1}\otimes\calD^{-}_{\lambda_2},\qquad
\nu=\lambda_1-\lambda_2
\end{equation}
with the generating function
\begin{equation}
\tilde{Y}^{+-\,\pm}_{\lambda_1,\lambda_2,\lambda}(z_1,z_2,w)
=\sqrt{\frac{\Gamma(\lambda\pm\nu)}{\Gamma(1-\lambda\pm\nu)}}\,
\biggl(\frac{z_1}{w}\biggr)^{\lambda_1}
\biggl(\frac{w}{z_2}\biggr)^{\lambda_2}
\frac{(1-z_1/z_2)^{\lambda-\lambda_1-\lambda_2}}
{(1-w/z_2)^{\lambda-\nu}(1-z_1/w)^{\lambda+\nu}}
\end{equation}
where the $\pm$ sign is linked to $c^{\pm}_{\lambda,m}$ in Eq.~(\ref{genfun_coeff}). Independent of that sign, the partial generating functions are
\begin{equation}
\tilde{Y}^{+-}_{\lambda_1,\lambda_2;\,\lambda,\,\nu+k}(z_1,z_2)
=z_1^{\lambda_1}z_2^{-\lambda_2}(1-z_1/z_2)^{-\lambda_1-\lambda_2}\,
\matelem{\mathring{\psi}}{\lambda}{\nu}{\nu+k}(z_1,z_2).
\end{equation}
where $\matelem{\mathring{\psi}}{\lambda}{\nu}{m}$ is defined by equation~(\ref{psi_symm}) with further reference to Eq.~(\ref{psi_lnm}).

The normalizability of the intertwiner $\Upsilon^{+-}_{\lambda_1,\lambda_2;\lambda}$ is related to the asymptotics of the partial generating function. We have $\tilde{Y}^{+-}_{\lambda_1,\lambda_2;\,\lambda,\,\nu+k}(z_1,z_2) =z_1^{\lambda_1+k}z_2^{-\lambda_2}f(z_1/z_2)$ with
\begin{equation}\label{psi_asymp}
f(z) \approx a_{+}(1-z)^{\alpha_+}+a_{-}(1-z)^{\alpha_-}\quad
\text{for }\, z\to 1,\qquad
\alpha_{+}=\lambda-\lambda_1-\lambda_2,\quad
\alpha_{-}=1-\lambda-\lambda_1-\lambda_2.
\end{equation}
The function $f$ is analytic in the complex plane with a branch cut from $1$ to $+\infty$. Its $n$-th Taylor coefficient $f_n$ can be expressed as a Cauchy integral over a circle of radius $r>1$ and around the branch cut section $[1,r]$, resulting in this asymptotic formula:
\begin{equation}
f_n\approx \frac{a_{+}\,n^{-1-\alpha_+}}{\Gamma(-\alpha_{+})}
+\frac{a_{-}\,n^{-1-\alpha_-}}{\Gamma(-\alpha_{-})}\quad
\text{for }\, n\to\infty.
\end{equation}
On the other hand, $c_{\lambda_1,m_1}^{+}\approx m_1^{\lambda_1-1/2}$ for $m_1\to+\infty$; similarly, $c_{\lambda_2,m_2}^{-}\approx (-1)^{m_2+\lambda_2}|m_2|^{\lambda_2-1/2}$ for $m_2\to-\infty$. Thus, equation~(\ref{psi_asymp}) translates to the following $n\to\infty$ asymptotics of the Clebsch-Gordan coefficients:
\begin{equation}
\bbra{\lambda_1+n+k,\,-\lambda_2-n} 
\Upsilon^{+-}_{\lambda_1,\lambda_2;\lambda}\bket{\nu+k}
\approx (-1)^{n}\biggl(
\frac{a_{+}\,n^{-\lambda}}{\Gamma(\lambda_1+\lambda_2-\lambda)}
+\frac{a_{-}\,n^{-1+\lambda}}
{\Gamma(\lambda_1+\lambda_2-1+\lambda)} \biggr).
\end{equation}
It is now easy to see that the vector $\Upsilon^{+-}_{\lambda_1,\lambda_2;\lambda}\ket{\nu+k}$ is normalizable or $\delta$-normalizable if the spinor $\matelem{\mathring{\psi}}{\lambda}{\nu}{\nu+k}$ is normalizable or $\delta$-normalizable. In addition, the said vector is normalizable if $\lambda_1+\lambda_2<\frac{1}{2}$ and $\lambda$ (or, equivalently, $1-\lambda$) is equal to $\lambda_1+\lambda_2$. This case corresponds to the complementary series representation $\calC^{\nu}_{\lambda(1-\lambda)}$.

To calculate the norm of the intertwiner for $\lambda=\frac{1}{2}+is$, we set $k=0$ so that the coefficients $a_{+}=a_{\lambda,\nu,-\nu}$ and $a_{-}=a_{1-\lambda,\nu,-\nu}$ are given by Eq.~(\ref{a_llr}). Thus,
\begin{equation}
\bigl(\Upsilon^{+-}_{\lambda_1,\lambda_2;\,1/2+is}\bigr)^{\dag}\,
\Upsilon^{+-}_{\lambda_1,\lambda_2;\,1/2+is'}
=\frac{\cosh(2\pi s)+\cos(2\pi\nu)}{2s\sinh(2\pi s)\,
|\Gamma(\lambda_1+\lambda_2-1/2-is)|^2}\,
\delta(s-s')\,\mathbf{1}^{\nu}_{1/2+is}.
\end{equation}
In the discrete series case $\calD^{+}_{\lambda}$ (with $\lambda=\lambda_1-\lambda_2-p>\frac{1}{2}$ and $p\in\{0,1,2,\ldots\}$), it is convenient to consider the Clebsch-Gordan coefficients with $m=\lambda$. We proceed with the calculation of the norm:
\useshortskip{\begin{gather}
\tilde{Y}^{+-}_{\lambda_1,\lambda_2;\lambda,\lambda}(z_1,z_2)
=(-1)^p\sqrt{\frac{\Gamma(2\lambda+p)}{\Gamma(2\lambda)\,p!}}\,
z_1^{\lambda_1}z_2^{-\lambda_2-p}(1-z_1/z_2)^{-2\lambda_2-p},
\displaybreak[0]\\[5pt]
\bbra{\lambda_1+r,\,-\lambda_2-p-r} 
\Upsilon^{+-}_{\lambda_1,\lambda_2;\lambda}\bket{\lambda_1-\lambda_2-p}
=(-1)^{r}
\sqrt{\frac{\Gamma(2\lambda+p)\,\Gamma(2\lambda_2+p+r)\,(p+r)!}
{\Gamma(2\lambda)\,\Gamma(2\lambda_2+p)^2\,\Gamma(2\lambda_1+r)\,p!\,r!}},
\displaybreak[0]\\[8pt]
(\Upsilon^{+-}_{\lambda_1,\lambda_2;\lambda})^{\dag}
\Upsilon^{+-}_{\lambda_1,\lambda_2;\lambda}
=\frac{1}{(2\lambda-1)\,\Gamma(\lambda_1+\lambda_2-\lambda)\,
\Gamma(\lambda_1+\lambda_2-1+\lambda)}\,
\mathbf{1}^{+}_{\lambda}.
\end{gather}}
In the special case $\lambda=\lambda_1+\lambda_2<\frac{1}{2}$, the partial generating function for $m=\nu$ is quite simple, namely $\tilde{Y}^{+-}_{\lambda_1,\lambda_2;\lambda,\nu}(z_1,z_2) =z_1^{\lambda_1}z_2^{\lambda_2}\, \hgfs(\lambda_1,\lambda_2,1;z_1/z_2)$. Summing up the squares of the Clebsch-Gordan coefficients gives this result:
\begin{equation}
(\Upsilon^{+-}_{\lambda_1,\lambda_2;\lambda})^{\dag}
\Upsilon^{+-}_{\lambda_1,\lambda_2;\lambda}
=\frac{\sin(2\pi\lambda_1)\,\sin(2\pi\lambda_2)\,\Gamma(1-2\lambda)}
{\pi^2}\,
\mathbf{1}^{\nu}_{\lambda}.
\end{equation}

In conclusion, the decomposition of identity is as follows:
\begin{equation}
\wideboxed{
\begin{aligned}
\mathbf{1}_{\lambda_1}^{+}\otimes\mathbf{1}_{\lambda_2}^{-}
&=\int_{0}^{\infty} ds\,
\frac{2s\,\sinh(2\pi s)\,
|\Gamma(\lambda_1+\lambda_2-1/2-is)|^2}
{\cosh(2\pi s)+\cos(2\pi\nu)}\,
\Pi^{+-}_{\lambda_1,\lambda_2;\,1/2+is}
\\[3pt]
&+\sum_{\substack{\lambda=|\nu|-p>1/2\\ p=0,1,2,\ldots}}
\bigl(2\lambda-1\bigr)\, 
\Gamma(\lambda_1+\lambda_2-\lambda)\,\Gamma(\lambda_1+\lambda_2-1+\lambda)\,
\Pi^{+-}_{\lambda_1,\lambda_2;\lambda}
\\[3pt]
&+\biggl(\frac{\pi^2}
{\sin(2\pi\lambda_1)\,\sin(2\pi\lambda_2)\,\Gamma(1-2\lambda)}\,
\Pi^{+-}_{\lambda_1,\lambda_2;\,\lambda}\quad
\text{for }\, \lambda=\lambda_1+\lambda_2<\frac{1}{2}
\biggr)
\end{aligned}
}
\end{equation}
where $\nu=\lambda_1-\lambda_2$. The expression for the projector is similar to that for spinors:
\begin{empheq}[box=\widebox]{gather}
\tilde{\Pi}^{+-}_{\lambda_1,\lambda_2;\lambda}(z_1,z_2;w_1,w_2)
=\frac{(z_1/w_1)^{\lambda_1}(w_2/z_2)^{\lambda_2}\,
(1-v)^{\lambda-\lambda_1-\lambda_2}}
{(1-z_1/w_1)^{2\lambda_1}(1-w_2/z_2)^{2\lambda_2}}\,
\hgfs\bigl(\lambda+\nu,\,\lambda-\nu,\,1;\,v\bigr)
\\[5pt]
v=u^{-1}=\frac{(z_1-w_2)(z_2-w_1)}{(z_1-w_1)(z_2-w_2)}
\end{empheq}

\section*{Acknowledgments}

I thank Josephine Suh for helpful comments as well as catching a number of errors in the paper draft. I gratefully acknowledge the support by the Simons Foundation under grant~376205 and through the ``It from Qubit'' program, and from the Institute of Quantum Information and Matter, a NSF Frontier center funded in part by the Gordon and Betty Moore Foundation.

\end{document}